\pdfoutput=1
\RequirePackage[normalem]{ulem}
\RequirePackage{color}\definecolor{RED}{rgb}{1,0,0}\definecolor{BLUE}{rgb}{0,0,1}

\def\isReadyToSubmit{0}   
\def\isCameraReady{1}     
\def\isAnonymousSubmission{0}  
\def\sandy{0}             
\def\confidential{0}      
\def\tightenTitles{1}     

\ifnum\sandy=0
    \newcommand{\subparagraph}{}
\documentclass[10pt,twocolumn]{IEEEtran}

\def\thesectiondis{\thesection.}                   
\def\thesubsectiondis{\thesectiondis\Alph{subsection}.}          
\def\thesubsubsectiondis{\thesubsectiondis\arabic{subsubsection}.} 

\else
  \documentclass[letterpaper,singlecolumn,10pt]{article}
\fi

\usepackage[top=1in, bottom=1in, left=1in, right=1in]{geometry}

\usepackage{times}
\usepackage{graphicx}
\usepackage{subfigure}
\usepackage{array} 
\usepackage{amsmath,amsfonts,amssymb,amsthm}
\usepackage{color}
\usepackage[hyphens]{url}
\usepackage{listings} 
\usepackage{xspace} 
\usepackage{algpseudocode} 
\usepackage{algorithm} 
\usepackage{wrapfig} 
\usepackage{flushend} 
\usepackage{multirow} 
\usepackage{booktabs} 
\usepackage{anyfontsize} 
\usepackage{paralist} 
\usepackage[font=small]{caption}
\usepackage{courier} 
\usepackage{verbatim}
\usepackage[usenames,dvipsnames,svgnames,table]{xcolor}
\usepackage{calc}
\usepackage{amsthm}
\usepackage{pbox}
\usepackage{algorithm, algpseudocode}
\ifnum\sandy=1
  \usepackage{setspace}
\fi

\ifnum\tightenTitles=1
    \usepackage[compact,small]{titlesec}
    \titlespacing{\section}{0pt}{*.6}{*.2}
    \titlespacing{\subsection}{0pt}{*.4}{*.2}
    \titlespacing{\subsubsection}{0pt}{*.4}{*.2}

    \usepackage{etoolbox}
    \makeatletter
    \patchcmd{\ttlh@hang}{\parindent\z@}{\parindent\z@\leavevmode}{}{}
    \patchcmd{\ttlh@hang}{\noindent}{}{}{}
    \makeatother

\fi

\let\oldenumerate\enumerate
\renewcommand{\enumerate}{
  \oldenumerate
  \setlength{\itemsep}{.5pt}
  \setlength{\parskip}{0pt}
  \setlength{\parsep}{0pt}
}

\let\olditemize\itemize
\renewcommand{\itemize}{
  \olditemize
  \setlength{\itemsep}{1pt}
  \setlength{\parskip}{0pt}
  \setlength{\parsep}{0pt}
}

\theoremstyle{definition}

\makeatletter
\renewcommand{\ALG@beginalgorithmic}{\footnotesize}
\makeatother

\newcommand{\eg}{{\em e.g.,~}}

\def\F{Fig.~}
\renewcommand{\figurename}{Fig.}

\newcommand{\ar}[3]{} 
\include{latexdefs}
\include{mathnotation}

\newcommand{\headingg}[1]{\noindent{\bf{#1}}} 
\newcommand{\heading}[1]{{\vspace{0pt}\noindent\bf{#1}}} 

{\makeatletter
 \gdef\xxxmark{%
   \expandafter\ifx\csname @mpargs\endcsname\relax 
     \expandafter\ifx\csname @captype\endcsname\relax 
       \marginpar{\textcolor{red}{xxx~}}
     \else
       \textcolor{red}{xxx~}
     \fi
   \else
     \textcolor{red}{xxx~}
   \fi}
 \gdef\xxx{\@ifnextchar[\xxx@lab\xxx@nolab}
 \long\gdef\xxx@lab[#1]#2{{\bf [\xxxmark \textcolor{red}{#2} ---{\sc #1}]}}
 \long\gdef\xxx@nolab#1{{\bf [\xxxmark \textcolor{red}{#1}]}}
 \ifnum\isReadyToSubmit=1
   \long\gdef\xxx@lab[#1]#2{}\long\gdef\xxx@nolab#1{}
 \fi
}

{\makeatletter
 \gdef\edit{\@ifnextchar[\edit@lab\edit@nolab}
 \long\gdef\edit@lab[#1]#2{[\textcolor{red}{#2} ---{\sc #1}]}
 \long\gdef\edit@nolab#1{[\textcolor{red}{#1}]}
 \ifnum\isReadyToSubmit=1
   \long\gdef\edit@lab[#1]#2{[#2]}
 \fi
}

\newcommand{\ignore}[1]{}

\definecolor{grey}{rgb}{0.5,0.5,0.5}

\newcommand{\code}[1]{\tt{#1}}

\newcommand{\x}{\vec{x}}        
\newcommand{\xprime}{\vec{x^\prime}}
\renewcommand{\l}{l}
\newcommand{\obs}{\langle\x,\l\rangle}
\newcommand{\obsprime}{\langle\xprime,\l\rangle}

\newcommand{\userId}{{\color{red} \textit{userId}}}   
\newcommand{\urlHash}{{\color{blue} \textit{urlHash}}}

\newcommand{\urlHashAdId}{\langle {\color{magenta} \textit{urlHash}, \textit{adId}} \rangle}

\usepackage{authblk}

\usepackage{times}
\usepackage{amsthm}
\usepackage[noadjust]{cite}

\ifnum\confidential=1
  \usepackage{fancyhdr}
\fi

\newcommand{\sysname}{Pyramid\xspace}

\IEEEoverridecommandlockouts
\ifnum\isAnonymousSubmission=0
  \author{\vspace{-0.2cm}Mathias Lecuyer$^{**1}$}
  \author{Riley Spahn$^{**1}$\thanks{$^{**}$First authors in alphabetical order.}}
  \author{Roxana Geambasu$^1$}
  \author{Tzu-Kuo Huang$^{\dagger2}$\thanks{$^{\dagger}$Work done while at
  Microsoft Research.}}
  \author{Siddhartha Sen$^3$}
  \affil{\vspace{-0.4cm}$^1$Columbia University, $^2$Uber Advanced Technologies Group, and $^3$Microsoft Research\vspace{-0.9cm}}
\else
  \author{Paper \# 183
  \vspace{-0.5cm}}
\fi

\begin{document}

\ifnum\sandy=1
  \doublespacing
\fi
\date{}

\title{
\vspace{-1cm}
{\huge \sysname: Enhancing Selectivity in Big Data Protection with Count
Featurization*\thanks{*Technical report version of the IEEE S\&P'17 paper with the same name and authors. 
This technical report describes a recent addition to \sysname to make some of our processes differentially private
(\S\ref{s:dp-processes}.}}
}

\maketitle

\ifnum\isCameraReady=1
  \thispagestyle{empty}
  \pagestyle{empty}
\fi

\setlength{\tabcolsep}{0.5em}

\begin{abstract}

Protecting vast quantities of data poses a daunting challenge for the growing
number of organizations that collect, stockpile, and monetize it.
The ability to distinguish data that is actually needed from data collected
``just in case'' would help these organizations to limit the latter's
exposure to attack.
A natural approach might be to monitor data use and retain only the working-set
of in-use data in accessible storage; unused data can be evicted to a highly
protected store.
However, many of today's big data applications rely on machine learning (ML)
workloads that are periodically retrained by accessing, and thus exposing to
attack, the entire data store.
Training set minimization methods, such as {\em count featurization},
are often used to limit the data needed to train ML workloads to improve
performance or scalability.

We present {\em \sysname}, a limited-exposure data
management system that builds upon count featurization to enhance data protection.
As such, \sysname uniquely introduces both the idea and proof-of-concept for
leveraging training set minimization methods to instill rigor and selectivity
into big data management.
We integrated \sysname into Spark Velox, a framework for ML-based targeting
and personalization.
We evaluate it on three applications and show
that \sysname approaches state-of-the-art
models while training on
less than {\em 1\% of the raw data}.

\end{abstract}

\section{Introduction}
\label{sec:introduction}

Driven by cheap storage and the immense perceived potential of ``big data,''
both public and private sectors are accumulating vast quantities of personal
data: clicks, locations, visited websites, social interactions, and more.
Data offers unique opportunities to improve personal and business
effectiveness.
It can boost applications' utility by personalizing their features; increase
business revenues via targeted product placement; improve social processes
such as healthcare, disaster response and crime prevention.
Its commercialization potential, whether real or perceived, drives
unprecedented efforts to grab and store raw data resources that can later be
mined for profit.

Unfortunately, this ``collect-everything'' mentality poses serious risks for
organizations by exposing extensive data stores to external and
internal attacks.
The hacking and exploiting of sensitive corporate and governmental information
have become commonplace~\cite{opm-hack, experian-data-breach}.
Privacy-transgressing employees have been discovered snooping into data stores
to spy on friends, family, and job candidates~\cite{nsa-employees-spy-on-exes,
celbs-medical-records}.
Although organizations strive to restrict access to particularly sensitive
data (such as passwords, SSNs, emails, banking data), properly
managing access controls for diverse and potentially sensitive
information remains an unanswered problem.

Compounding this challenge is a significant new thrust in the public and
private spheres to integrate data collected from multiple sources into a
single, giant repository (or ``data lake'') and make that available to any
applications that might benefit from it~\cite{hearst-data, google-data-policy,
azure-data-lake}. This practice magnifies the data exposure problem,
transforming big data into what some have called a ``toxic asset''~\cite{data-toxic-asset}.

Our goal in this paper is to explore a more rigorous and selective approach to
big data protection.
We hypothesize that not all data that is collected and archived is, or may
ever be, needed or used.
The ability to distinguish data needed now or in the future from data
collected ``just in case'' could enable organizations to restrict the latter's
exposure to attacks.
For example, one could ship unused data to a tightly controlled store, whose
read accesses are carefully mediated and audited. Turning this hypothesis into a
reality requires finding ways to: (1) minimize data kept in the company's
widely-accessible data lakes, and (2) avoid the need to access the controlled
store to meet current and evolving workload needs.

A natural approach might be to monitor data use and retain only the working set
of in-use data in accessible storage; data unused for some time is evicted to the
protected store~\cite{cleanos}.
However, many of today's big data applications involve machine learning (ML)
workloads that are periodically retrained to incorporate new data,
resulting in frequent accesses to all data.
How can we determine and minimize the training set---the ``working set'' for
emerging ML workloads---to adopt a more rigorous and selective approach to
big data protection?

We observe that for ML workloads, significant research is devoted to limiting
the amount of data required for training.
The reasons are many but typically do not involve data protection.
Rather, they include increasing performance, dealing with sparsity, and
limiting labeling effort.
Techniques such as dimensionality reduction~\cite{burges2010dimension},
feature hashing~\cite{shi2009hash}, vector quantization~\cite{gersho2012vector},
and count featurization~\cite{export:260043} are
routinely applied in practice to reduce data dimensionality so models can be
trained on manageable training sets.
Semi-supervised~\cite{Zhu06semi-supervisedlearning} and active learning~\cite{settles2012active} reduce
the amount of labeled data needed for training when labeling requires manual
effort.

{\em Can such mechanisms also be used to limit exposure of the data being
collected?
How can an organization that already uses these methods develop a more
robust data protection architecture around them?
What kinds of protection guarantees can this architecture provide?}

As a first step to answering these questions, we present {\em \sysname}, a
limited-exposure big-data management system built around a specific
training set minimization method called {\em count
featurization}~\cite{Chapelle:2014:SSR:2699158.2532128, Chen:2009:LBT:1557019.1557048, conf-kdd-LiWZCMJ10, export:260043}.
Also called historical statistics, count featurization is a 
widely used technique for reducing training times by feeding ML algorithms
with a limited subset of the collected data combined (or {\em
featurized}) with historical aggregates from much larger amounts of
data.
The method is valuable when features with strong predictive power are highly
dimensional, requiring large quantities of data (and large
amounts of time and resources) to be properly modeled.
Applications that use count featurization include targeted
advertising, recommender systems, and content personalization systems.
Such applications rely on user information to predict clicks, but since
there can be hundreds of millions of users, training can be very 
expensive without some way to aggregate users, like
count featurization.
The advertising systems at Microsoft, Facebook, and Yahoo are all built upon
this 
mechanism~\cite{personalCommunicationWithMisha},
and Microsoft Azure offers it as a service~\cite{dracula}.

\sysname builds on count featurization to construct a selective data
protection architecture that minimizes exposure of individual observations
(e.g., individual clicks).
To highlight, \sysname: keeps a small, rolling window of accessible raw data (the {\em hot window}); summarizes the history with privacy-preserving aggregates (called {\em counts}); 
trains application models with hot raw data featurized with counts; and
rolls over the counts to forget all traces of observations past a specified
retention period.
Counts are infused with differentially private
noise~\cite{Dwork:2006:CNS:2180286.2180305} to protect individual observations
that are no longer in the hot window but still fall within the retention
period.
Counts can support modifications and additions of many (but not all) types of
models; historical raw data, which may be needed for workloads not supported
by count featurization, is kept in an encrypted store whose decryption requires special access.

While count featurization is not new, our paper is the first to
retrofit it for data protection.
Doing so raises significant challenges. We first need to define meaningful requirements and protection guarantees that can be achieved with this mechanism, such as the amount of exposed information or the granularity of protection.
We then need to achieve these protection guarantees without affecting model accuracy and scalability, despite using much less raw data.
Finally, to make the historical raw data store easier to protect, we need to access it as little as possible. This means supporting workload evolution, such as parameter tuning or trying new algorithms, without the need to go back to historical raw data store.

We overcome these challenges with three main techniques: (1) {\em weighted
noise infusion}, which automatically shares the privacy budget to give 
noise-sensitive features less noise; (2) an {\em unbiased private count-median sketch}, a data structure akin to a count-min
sketch that resolves the large negative bias arising from   
applying differentially private noise to a count-min sketch; and (3) {\em automatic
count selection}, which detects potentially useful groups of features to count
together, to avoid accesses to the historical data. Together, these
techniques reduce the impact of differentially private noise and count
featurization.

We built \sysname and integrated it into Spark Velox, a targeting and
personalization framework, to add rigor and selectivity to its data
management.
We evaluated three applications: a targeted advertising system
using the Criteo dataset, a movie recommender using the MovieLens
dataset, and MSN's production news personalization system.
Results show that: (1) Pyramid approaches
state-of-the-art models while training on less than {\em 1\% of the raw data}.
(2) Protecting historical counts with differential privacy has only {\em
2\% impact on accuracy}.
(3) Pyramid adds
just {\em 5\% performance overhead}.

Overall, we make the following contributions:

\begin{enumerate}
\item Formulating the {\em selective data protection
problem} for emerging ML workloads as a training set minimization problem, for which many
mechanisms already exist. 

\item The design of \sysname, the first selective data management system
that minimizes data exposure in anticipation of attack.
Built upon 
count featurization, \sysname is particularly suited for targeting and
personalization workloads.

\item A set of new techniques to balance solid protection
guarantees with model accuracy and scalability, such as our
unbiased private count-median sketches.

\item \sysname's code, both integrated into Spark Velox and as a stand-alone
library ready to integrate in other targeting/personalization frameworks.
{\tt \url{https://columbia.github.io/selective-data-systems/}}
\end{enumerate}

\section{Motivation and Goals}
\label{sec:motivation}

This paper argues for needs-based selectivity in big data protection:
protecting data differently depending on whether or not it is actually needed to
handle a company's day-to-day workloads. 
Intuitively, data that is needed day-to-day is less amenable to certain kinds
of protection (e.g., auditing or case-by-case access control) than
data needed only for exceptional situations.
A key question is {\em whether a company's day-to-day needs can be captured
with a limited and well-defined data subset}.
While we do not claim to answer this question in full, we present with
\sysname the first evidence that selectivity can be achieved in one 
important big-data workload domain: {\em ML-based targeting and
personalization}.
The following scenario motivates selectivity and shows how and in what contexts
\sysname helps improve protection.

\subsection{Example Use Case}
\label{sec:example-scenario}

MediaCo, a media conglomerate, collects {\em observations} of user
behavior from its hundreds of affiliate news and entertainment sites.
Observations include the articles users read and share, the ads they click,
and how they respond to A/B testing. MediaCo uses this data to optimize various
processes, including recommending articles to users, showing the most relevant
articles first, and targeting ads. Initially, MediaCo collected observations
from affiliate sites in separate, isolated repositories; different engineering teams used different repos to optimize these processes for each affiliate site.
Recently, MediaCo has started to track users across sites using cookies and to
integrate all data into a central data lake.  Excited about the
potential of the much richer information in the data lake, MediaCo plans to
provide indiscriminate access to all engineers. However, aware of recent
external hacking and insider attacks affecting other companies, it worries about
the risks it assumes with such wide access.

MediaCo decides to use \sysname to limit the exposure of historical
observations in anticipation of such attacks.  
For MediaCo's {\em main workloads}, which consist of targeting and
personalization, the company already uses count featurization to address
sparsity challenges; hence, \sysname is directly applicable for those
workloads.
They configure it by keeping \sysname's hot window of raw observations, along
with its noise-infused historical statistics, in the widely accessible data
lake so all engineers can train their models, tune them, and explore new
algorithms every day.
\sysname absorbs many workload needs---current and evolving---as long as the
algorithms draw on the same user data to predict the same outcome (e.g.,
whether a user will click on an ad).
MediaCo also configures a one-year retention period for all observations;
after this period, \sysname removes observations from the statistics and
launches retraining of all application models to purge the old activity.
Finally, MediaCo stores all raw observations in an encrypted store whose
read accesses are disabled by default.
Access to this store is granted temporarily and on a case-by-case basis to
engineers who demonstrate the need for statistics beyond those that \sysname
maintains.

In addition to targeting/personalization workloads, MediaCo has {\em
other, potentially non-ML workloads}, such as business analytics, trend
studies, and forensics; for these, count featurization may not apply.
Hence, MediaCo gives direct access to the raw-data store to engineers
managing these workloads and isolates their computational resources from the
targeting/personalization teams.

With this configuration, MediaCo minimizes access to its collected data on a
needs basis.
Assuming no entity with full access to the historical raw data is malicious,
\sysname guarantees the following (detailed in \S\ref{sec:threat-model}).
(1) Any observations preceding the hot window when an attack begins will
be hidden from the attacker.
(2) Hiding is done at an individual observation level during the retention
period and in bulk past the retention period.
(3) Only in exceptional circumstances do engineers get access to the
historical raw data.
With these guarantees, MediaCo negotiates lower data loss insurance
premiums and gains PR benefits for its efforts to protect user data.

\subsection{Threat Model}
\label{sec:threat-model}

\begin{figure}[t]
  \centering
  \footnotesize
  \includegraphics[width=0.95\linewidth]{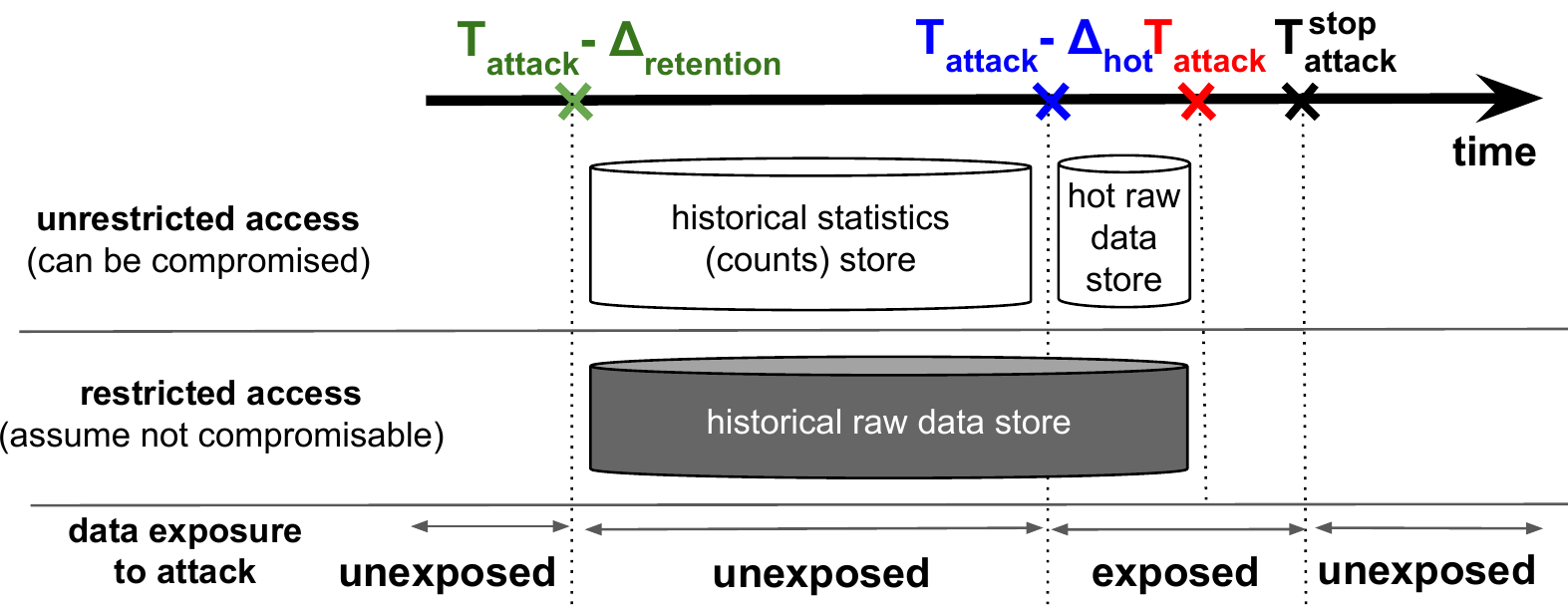}
  \vspace{-0.2cm}
  \caption{{\bf Threat model.}
      $T_{\textit{attack}}$: time the attack starts; $T_{\textit{attack}}^{\textit{stop}}$: time the attack
      is eradicated; $\Delta_{\textit{hot}}$: hot window length; $\Delta_{\textit{retention}}$: company's
    data retention period.
  }  
  \label{fig:threat-model}
  \vspace{-0.5cm}
\end{figure}

\F~\ref{fig:threat-model} illustrates \sysname's threat model and guarantees.
\sysname gives guarantees similar to those of forward secrecy: a one time compromise will not allow an adversary to access all past data.
Attacks are assumed to have a well-defined start time, $T_{\textit{attack}}$, when the adversary gains access to the machines charged with running \sysname, and a well-defined end time, $T_{\textit{attack}}^{\textit{stop}}$, when administrators discover and stop the intrusion.
Adversaries are assumed to not have had access to the system before
$T_{\textit{attack}}$, nor to have performed any action in anticipation of their attack
(e.g., monitoring external predictions, the hot window, or the
models' state), nor to have continued access after $T_{\textit{attack}}^{\textit{stop}}$.
The attacker's goal is to exfiltrate individual observations of user activities
(e.g., to know if a user clicked on a specific article/ad).
Historical raw data is assumed to be protected through independent means and not
compromised in this attack. \sysname's goal is to limit the hot data
in active use, which is widely accessible to the attacker.

Examples of adversaries that fit our threat model can be found among both the internal and external adversaries of a company.
An external adversary may be a hacker who breaks into the company's computing infrastructure at time $T_{\textit{attack}}$ and starts looking for data that may prove of value (e.g., information about celebrities' specific activities, what they liked or disliked, where they were in the past, etc.).
An internal adversary may be a privacy-transgressing employee who {\em spontaneously} decides at $T_{\textit{attack}}$ to look into some past action of a family member or friend (e.g., to check if the person has visited or liked a particular page).

After compromising \sysname's internal state, the attacker will gain access to data in three different representations: the hot data store containing plaintext observations, the historical counts, and the trained models themselves.
The plaintext observations in the hot data store are not protected in any way.
The historical statistics store contains differentially private count tables of
the recent past.  The attacker will learn some information from the count tables
but individual records will be protected with a differentially private
guarantee.  \sysname forces models to be retrained when observations are removed
from the hot raw data store, so the attacker will not be able to learn
anything from the models beyond what they have already learned above.

\sysname provides three protection levels:
\vspace{-0.1cm}

\begin{enumerate}
\item[{\bf P1}] {\em No protection for present or future observations}.
Observations in the hot data store when the attack begins, plus observations
added to the hot data store while the attack is ongoing, receive no
protection; i.e., observations received between 
($T_{\textit{attack}}-\Delta_{\textit{hot}}$) and 
$T_{\textit{attack}}^{\textit{stop}}$ receive no protection.

\item[{\bf P2}] {\em Protection for individual observations for the length of the retention period}.
Statistics about observations are retained in differentially private count
tables for a predefined retention period 
$\Delta_{\textit{retention}}$. The attacker may learn broad statistics about observations in
the interval $[T_{\textit{attack}}-\Delta_{\textit{retention}}, T_{\textit{attack}}-\Delta_{\textit{hot}}]$ but will not be able to confidently determine if a specific observation is present in the table.

\item[{\bf P3}] {\em Protection in bulk past the retention period}. Observations
    past their retention period (i.e.,  older than $T_{\textit{attack}}-\Delta_{\textit{retention}}$) have been phased out of the
historical statistics store and are protected separately by the historical raw data store.

\end{enumerate}
\vspace{-0.1cm}

Finally, we assume that no states created based on the hot raw data persist once the hot window is
rolled over.
While we explicitly launch retraining of models registered with \sysname, we operate under the assumption
that (1) the models' states are securely erased~\cite{securedelete:92} and (2) no other state was created out of band
based on the raw hot data (such as copies made by programmers).

\subsection{Design Requirements}
\label{sec:requirements}

Given the threat model, our design requirements are:
\vspace{-0.1cm}
\begin{enumerate}
\item[{\bf R1}] {\em Limit widely accessible data.}  
The hot data window is exposed to attackers; hence, \sysname must limit its
size subject to application-level requirements, such as the accuracy of models
trained with it.

\item[{\bf R2}] {\em Avoid accesses to historical raw data even for evolving workloads.}
\sysname must absorb as many current and evolving workload needs as possible
to limit access to the historical raw data. 

\item[{\bf R3}] {\em Support retention policies.}
\sysname must enforce a company's retention policies.
Although \sysname provides a differential privacy guarantee,
no protection is stronger than securely deleting data.

\item[{\bf R4}] {\em Limit impact on accuracy, performance, scalability.}
We intend to preserve the functional properties of applications and models
running on \sysname.

\end{enumerate}

\section{The \sysname Architecture}
\label{sec:design}

\sysname, the first selective data management architecture, builds upon
the ML technique of count-based featurization and augments it with new
mechanisms to meet the preceding design requirements.

\subsection{Background on Count-Based Featurization}
\label{sec:background}

Training predictive models can be challenging on data that contains
categorical variables (features) with large numbers of possible values (e.g.,
an ID or an interest vector).
Existing ML techniques that handle large feature spaces often make
strong assumptions about the data, e.g., assuming a linear relationship between the
features and the label (e.g., Lasso~\cite{tibshirani1996regression}).
If the data does not meet these assumptions, results can be very poor.

Count-based featurization~\cite{export:260043} is a popular
approach to handling categorical variables of high cardinality.
Rather than directly using the value of a categorical variable, this technique
featurizes the data with the number of times a particular feature value (e.g.,
a user ID) was observed with each label and the conditional
probability of the label given the feature value.
This substantially reduces dimensionality.
Suppose the raw data contains $d$ categorical features with an average
cardinality of $K$ and a label of cardinality $L$, where $K \gg L$;
e.g., in click prediction $K$ can be millions (number of users),
while $L$ is 2 (click, non-click).
Standard encoding of categorical variables~\cite{agresti2013categorical} results in
a feature space of dimension $O(d K)$, whereas with count featurization it is
$O(d L)$. Count featurization can also be applied to continuous variables or
continuous labels by first discretizing them; this increases 
dimensionality but only by a small factor.

The dramatic dimensionality reduction yields important
benefits.
It is known that fewer dimensions permit more efficient learning, both
statistically and computationally, potentially at the cost of reducing
predictive accuracy. However, count featurization makes it feasible to apply
advanced, nonlinear models, such as neural networks, boosted trees, and random forests.
This combination of succinct data representation and powerful learning
models enables substantial reduction of the training data with little
loss in predictive performance.
Quantified in \S\ref{sec:evaluation}, this is the insight behind our use of
count-based featurization to limit data exposure.

\begin{figure}[t]
  \centering
  \footnotesize
  \includegraphics[width=\linewidth]{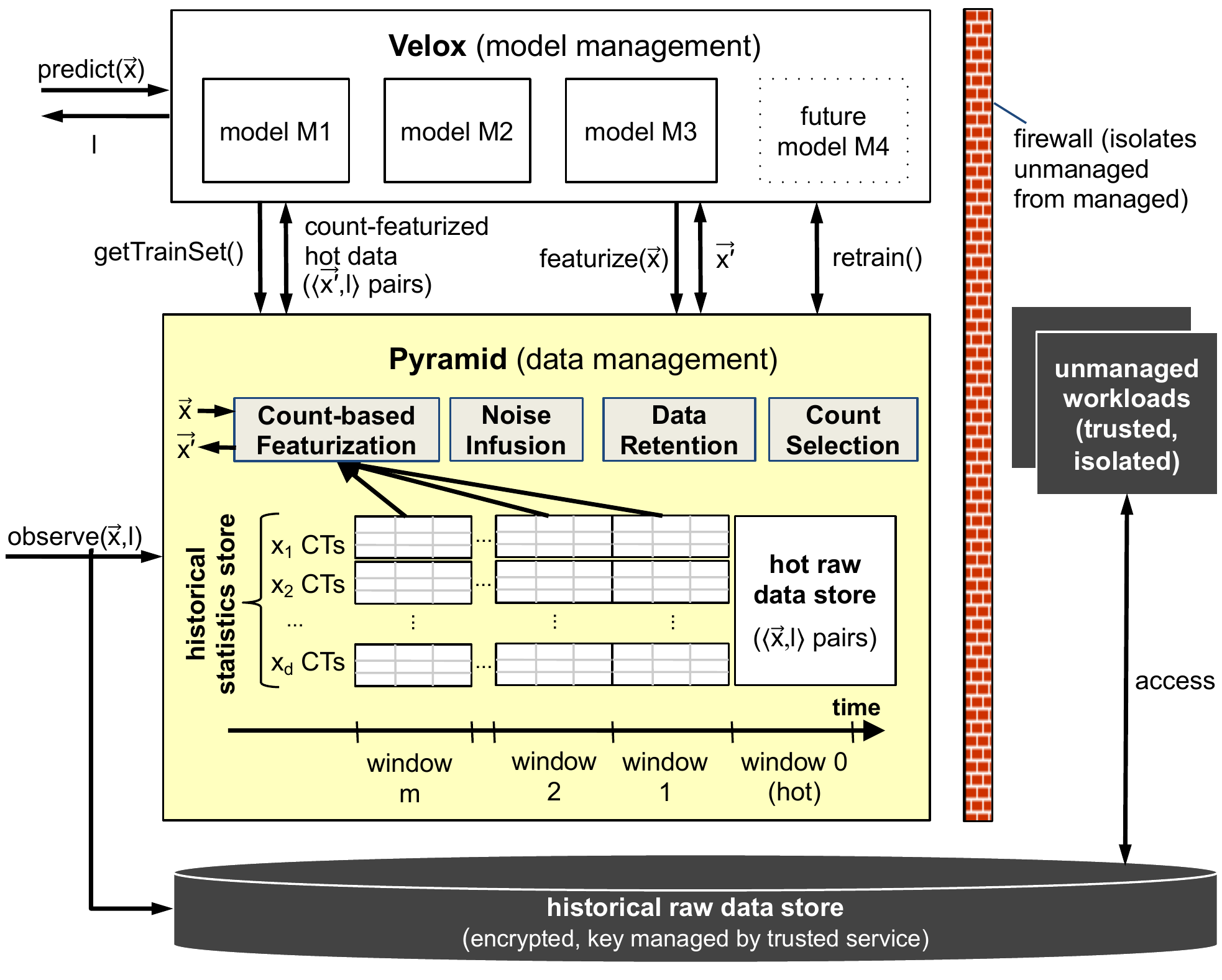}
  \vspace{-0.5cm}
  \caption{
    {\bf \sysname's architecture.} Notation:
    $\x$:  feature vector;
    $\l$: label; $\xprime$: count-featurized feature vector; CT: count table.
  }  
  \label{fig:architecture}
\vspace{-0.5cm}
\end{figure}

\subsection{Architectural Components}
\label{sec:overview}

\F\ref{fig:architecture} shows \sysname's architecture.
\sysname manages collected data (observations) on behalf of application
models hosted by a model management system.
In our case, we use Velox~\cite{velox}, built on Spark.
Velox facilitates ML-based targeting and personalization services by
implementing three functions:
(1) fast, but incomplete, incorporation of new observations into models
that programmers register with Velox;
(2) low-latency prediction serving from these models; and
(3) periodic retraining of the models to correct inconsistencies created by
the incomplete incorporation of new observations.
Velox saves observations in a separate data management component, Spark's
Tachyon.
\sysname replaces this component to ensure rigorous and selective protection
of observations.

\sysname itself consists of four architectural components, shown across the top
of the highlighted box in \F\ref{fig:architecture}.
The first is {\em count featurization}, which leverages the known
ML mechanism to count featurize observations before feeding them to models for
training and prediction.
The second, third, and fourth are {\em noise infusion}, {\em data retention},
and {\em count selection}, which augment count featurization with
differential privacy and a set of new mechanisms to meet \sysname's design
requirements.
We discuss each component in turn.

\subsubsection{Count Featurization}
\label{sec:featurization-example}

\sysname hijacks the stream of observations collected by Velox (the
{\small \code observe} method) and count-featurizes them.
An observation is a pair $\obs$ with a feature vector $\x=\langle x_1, x_2,
..., x_d \rangle $ and a label $\l$.
Application models predict the label (or a probability for each possible label)
for a given feature vector by training on count-featurized observations.
When an observation arrives, \sysname incorporates it into two data
structures: (1) the {\em hot raw data store}, which retains observations from
the recent past, and (2) the {\em historical statistics store}, which consists
of multiple {\em count tables} that maintain the number of occurrences of each
feature with each label.
We maintain count tables for all features in $\x$ and for some feature combinations.
A separate set of count tables is maintained for each time window.

Featurization transforms a feature vector $\x$ into a count-featurized
feature vector $\xprime$, by replacing each feature $x_i$
with the conditional probabilities of each label value given $x_i$'s value.
The conditional probabilities are computed directly from the count tables as
discussed below.
To train its models, an application requests a training set from \sysname
({\small \code getTrainSet}).
\sysname featurizes the hot raw data with historical counts and returns it
to the application.
To predict the label for a feature vector $\x$, the application requests its
featurization from \sysname ({\small \code featurize}); \sysname returns
$\xprime$.

\begin{figure}[t]
  \centering
  \footnotesize
  \includegraphics[width=0.77\linewidth]{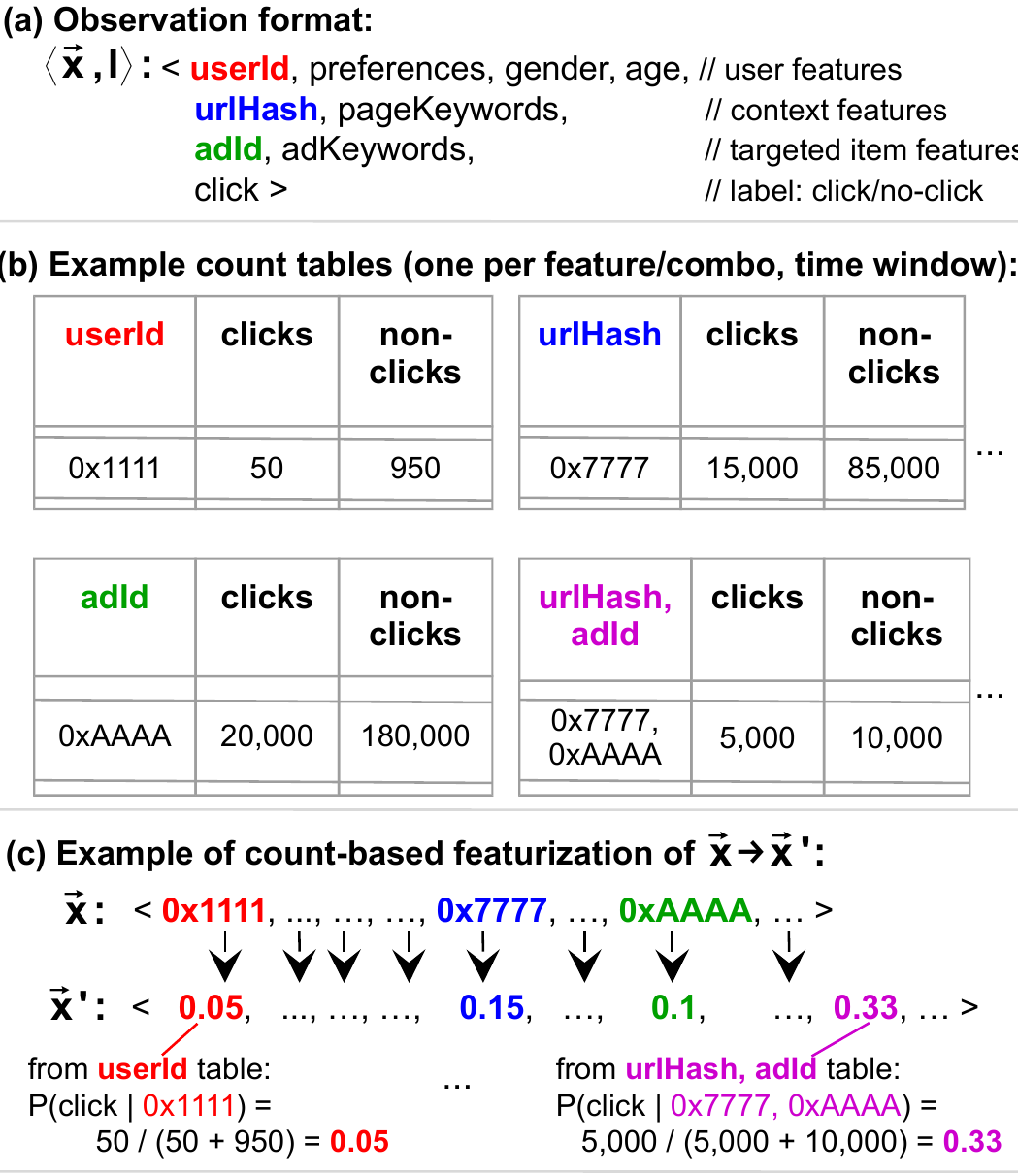}
  \vspace{-0.2cm}
  \caption{
    {\bf Count featurization example.}
  }  
  \label{fig:count-featurization-example}
\vspace{-0.5cm}
\end{figure}

\heading{Example.}
\F\ref{fig:count-featurization-example} shows (a) a sample observation format,
(b) some count tables used by \sysname to count-featurize it, and (c) a
sample count-featurized observation.

\noindent
$\bullet$
{\em Observation format.}
In targeting and personalization, an observation's feature vector
$\x$ typically consists of {\em user features} (e.g., id, gender, age, and
previously compiled preferences) and {\em contextual information} for the
observation (e.g., the URL of the article or the ad shown to the user, plus any
features of these).
The label $\l$ might indicate whether the user clicked on the article/ad.

\noindent
$\bullet$
{\em Count tables.}
Once an observation stream of the preceding type is registered with \sysname,
the $\userId$ table maintains for each user the number of clicks
the user has made on any ad shown and the number of non-clicks; it
therefore encodes each user's propensity to click on ads.
The $\urlHash$ table maintains for each URL the number of clicks
that each user made on any ad shown on that page; it therefore encodes the
page's inherent ``ad-clickability.''
\sysname maintains count tables for every feature in $\x$ and for some
feature combinations with predictive potential, such as the
$\urlHashAdId$ table, which encodes the
joint probability of a particular ad being clicked when it is shown on a
particular page.

\noindent
$\bullet$
{\em Count featurization.}
To count-featurize a feature vector {\small $\x = \langle x_1, x_2, \ldots,
x_d \rangle$},
\sysname first replaces each of its features with the conditional
probabilities computed from the count tables, e.g.,
{\small $\xprime = \langle P(\textit{click}|x_1), P(\textit{click}|x_2), \ldots,
P(\textit{click}|x_d) \rangle$}, where
{\small $P(\textit{click}|x_i) = \frac{\textit{clicks}}{\textit{clicks} + \textit{non\mbox{-}clicks}}$} from the row
matching the value of $x_i$ in the table corresponding to $x_i$.
\sysname also appends to $\xprime$ the conditional probabilities for any
feature combinations it maintains.
\F~\ref{fig:count-featurization-example}(c) shows an example of
feature vector $\x$ and its count-featurized version $\xprime$.
This is a simplified version of the count featurization function.  We can also
include the raw counts in $\xprime$, and support non-binary categorical labels
by including conditional probabilities for each label.
To avoid featurizing with an effectively random probability when a given
feature value has very few counts, we estimate the variance of our probability
estimate and, if it is too high, featurize with a default probability {\small
$P(click)$}.

\noindent
$\bullet$
{\em Training and prediction.}
Suppose a boosted-tree model is trained on a count-featurized dataset
($\obsprime$ pairs).
It might find that for users with a click propensity over {\color{red} 0.04},
the chances of a click are high for ads whose clickability exceeds
{\color{ForestGreen} 0.05} placed on websites with ad-clickability over
{\color{blue} 0.1}.
In this case, the model would predict a ``click'' label for the feature vector
in \F\ref{fig:count-featurization-example}(c).

\heading{Process.}
\sysname count-featurizes all features $x_i$ for each observation type.
For categorical features, we featurize them as described above.
For low-cardinality features, we can additionally include the raw feature values
in $\xprime$ alongside the conditional probabilities.
Continuous features are first mapped to a discrete space, binning them by
percentiles, and then count-featurized as categorical.
We do the same with continuous labels.

\sysname maintains hot windows and count tables as follows.
There is one hot window for each observation stream.
There is one count table per feature or feature group; it has 
a column for each label and a row for each value the feature can take.
To support granular retention times, each count table is composed of 
multiple windowed count tables holding data for observations collected during 
disjoint windows of time.
The complete count table is the sum of the associated windowed count tables.
When a new observation arrives, it is added to the hot window and made
immediately available to the models for (re)training.
The hot window is a sliding window that may be sized differently from the 
count table window.
It is also added to the current windowed count table; this count table
is withheld when computing the complete count table until it is finished
populating. At this point, \sysname begins using it
as part of the featurization process, phases out the oldest count table if it is
past its retention period, and begins populating a new count table
that has been initialized with differentially private noise.
Once count tables are incorporated into the featurization process, they are
never updated again.

\heading{Count-min sketches (CMSes).}
A key challenge with count featurization is its storage requirement.
For a categorical variable of cardinality $K$ and a label of cardinality $L$,
the count table is of size $O(LK)$.
A common solution, used in Azure~\cite{dracula}, is to store each table in
a Count-Min Sketch (CMS)~\cite{cormode2005cms}, a data structure that approximates counts in sub-linear space.
A CMS consists of a 2D array with an independent hash function for each row.
When a new feature arrives, the CMS uses the hash function for each row to assign the feature to a column and increment the value in that cell.

We query the CMS for a feature count by hashing the feature into a column of each row and taking the minimum value.
Despite overcounting from collisions, 
CMS provides sufficiently accurate count estimates to train ML models.
With a CMS, we can maintain more and/or larger count tables with bounded storage overheads.
This gives developers flexibility in the types of modeling they can do atop in-use data without tapping into the historical data store.
The CMS poses challenges to our noise infusion process, as described next.

\subsubsection{Noise Infusion}
\label{sec:differential-privacy}

\sysname's key contribution is to retrofit count featurization, a
technique developed for performance and scalability, to protect
past observations against exposure to attack.
\sysname infuses noise into the count tables to protect these observations.
While we leverage differential privacy methods~\cite{Dwork:2006:CNS:2180286.2180305}, correctly
applying these methods in our context poses scaling challenges.
For example, each observation contributes to multiple count tables, increasing
the noise required to guarantee differential privacy, and a na\"ive application
degrades accuracy when there are many count tables.
We present two techniques to address this challenge.
First, we use a {\em weighted noise infusion} technique to mitigate the
impact of noise, allowing us to navigate the privacy/utility trade-off.
Second, for high noise levels, we replace the CMS by a count-median sketch
\cite{Charikar:2002:FFI:646255.684566}, a data structure with weaker accuracy
guarantees than CMS but that provides an unbiased frequency estimate, making it
more robust to negative noise values.
To our knowledge, we are the first to observe that the count-median sketch structure is
better suited to differential privacy.
After a brief overview of differential privacy, we describe these techniques.

\heading{Differential privacy properties.}
\sysname's noise infusion component uses four differential privacy properties:

{\em 1. Privacy guarantees:}
Let $D_1$ be the database of past observations, $D_2$ be a database that differs
from $D_1$ by exactly one observation (i.e., $D_2$ adds or removes 1 observation), and $S$ the range of all possible count tables that can result from a randomized query $Q()$ that builds a count table from a window of observations.
The count table query $Q()$ is {\em $\epsilon$-differentially private} if  {\small $P[Q(D_1) \in S] \leq e^{\epsilon} \times P[Q(D_2) \in S]$}.
In other words, adding or removing an observation in $D_1$ does not
significantly change the probability distribution of possible count tables; therefore, the count table does not leak significant information about any specific observation~\cite{Dwork:2006:CNS:2180286.2180305}.
$\epsilon$ is called the query's {\em privacy budget}.

{\em 2. Laplace distribution:}
Let a query's {\em sensitivity} be the magnitude of the change in the query result triggered by adding or removing a single observation.
If the query has sensitivity $\Delta$, 
then adding noise drawn from a Laplace distribution with scale parameter $\frac{\Delta}{\epsilon}$ guarantees that the result is $\epsilon$-differentially private~\cite{Dwork:2006:CNS:2180286.2180305}.
Increasing $\frac{\Delta}{\epsilon}$ increases the standard deviation of the
distribution (stdev of a Laplace distribution with parameter $b$ is $b\sqrt{2}$).

{\em 3. Composability:} Differentially private queries are composable:
the sum of $n$ $\epsilon_n$-differentially private queries
is $(\Sigma\epsilon_n)$-differentially private~\cite{Mcsherry:pinq}.
This lets us maintain multiple count tables, possibly with different
budgets, and combine them without breaking guarantees. (Advanced composition
theorems allow sublinear loss in the privacy budget by relaxing the
guarantees to $(\epsilon, \delta)$-differential privacy~\cite{privacybook}, but
we do not explore that here.)

{\em 4. Post-processing resilience:} Any computation on a differentially
private data release remains differentially private~\cite{privacybook}. 
This is a crucial point for \sysname's protection guarantees: it ensures that guarantee {\bf P2}, the protection of individual past observations during their lifetime, holds for each model's internal state and outputs.
As long as models comply with retrain calls and erase all internal state when they do, their output is differentially private with regard to observations outside the hot window.

\heading{Basic noise infusion process.}
We apply these known properties when creating count tables for the hot
window.
Upon creating a count table, we initialize each
cell of the CMS storing that table with a random draw from a Laplace distribution.
This noise is added only once: the count tables are updated as observations
arrive and are sealed when the hot window rolls over.
To determine the correct parameter for the Laplace distribution, $b$, we must
account for three factors: (1) the internal structure of the CMS, (2) the
number of observations we want to hide simultaneously, and (3) the number of
count tables (features or feature combinations) we are maintaining.

First, an exact count table has sensitivity $1$ since adding or removing an
observation can only change one count by 1.
For a CMS, each observation is counted once per hash function; hence, the
sensitivity is $h$, the number of hash functions.
Second, if we aim to hide any group of $k$ observations with a privacy budget
of $\epsilon$, then we make a count table $\epsilon$-differentially private by
adding noise from a Laplace distribution of parameter $b=\frac{hk}{\epsilon}$
in every cell of the CMS.  
Third, we must maintain multiple count tables for the different features and
feature groups. Since each observation affects every count table, we
need to split the privacy budget $\epsilon$ among them, e.g., splitting
it evenly by adding noise with $b=\frac{nhk}{\epsilon}$ to each table.

The third consideration poses a significant challenge for \sysname: the amount
of noise we apply grows linearly with the number of count tables we keep.
Since the amount of noise directly affects application accuracy, this yields a
protection/accuracy tradeoff, which we address with weighted noise infusion.

\heading{Weighted noise infusion process.}
We note that count tables are not all equally susceptible to noise.
For example in our movie recommender, the $user$ table most likely contains low values, since each user rates only a few movies ($29$ for the median user).
Moreover, we do not expect this count to change significantly when adding more data, since single users will not rate significantly more movies.
Each $genre$ table however contains higher values (1M or more), since each genre characterizes multiple movies, each rated by many users.
Sharing noise equally between tables would pollute all counts by a standard
deviation of $145$ ($\epsilon=1$, $h=5$, and $k=1$), a reasonable amount for
$genre$s, but devastating for the $user$ feature, which essentially becomes
random.

\sysname's weighted noise infusion distributes the privacy budget unevenly across count tables, adding less noise to low-count features.
This way, we retain more utility from those tables, and the composability property of differential privacy preserves our protection guarantees.
Each table's share of noise is determined automatically, based on the count values observed in the hot window.
Specifically, the user specifies a quantile, and the privacy budget is shared
between each feature proportionally to this quantile of its counts.
For instance we use the first percentile, so that 99\% of the counts for a feature will be less affected by the noise.
Sharing the privacy budget proportionally to the counts is a heuristic that
makes the noise's standard-deviation proportional to the typical counts of each feature.
This scheme is also independent of the learning algorithm.

Section~\ref{sec:evaluation} shows that weighted noise infusion is
vital for providing protection while preserving accuracy at scale: without it, the cost of hiding
single observations is a 15\% accuracy loss; with it, the loss is less than 5\%.

The weight selection process must be made differentially private lest it
may leak information about the hot window used to compute the weights.
While our IEEE Security \& Privacy paper~\cite{pyramid-sp17} did not address
this problem, we have since modified \sysname to compute feature weights in a
differentially private way. 
\S\ref{s:dp-processes} describes our method, which can be summarized as follows.
We compute the weights every so often (e.g., every month) using the data in one hot window.
We use a configurable portion of one window's privacy budget and leverage smooth
sensitivity~\cite{Nissim:2007:SSS:1250790.1250803} to compute differentially private
count percentiles, which we then use as feature weights.
We compute differentially private percentiles by adapting the J-List algorithm for the differentially private median described in~\cite{Nissim:2007:SSS:1250790.1250803}.
\S\ref{s:dp-weights} shows that we can make the weighted noise infusion calculation
differentially private without reducing the accuracy wins gained from doing weighted noise
infusion.

\heading{Unbiased private count-median sketch.}
Another factor that degrades performance when adding differentially private noise is the interaction between the noise and the CMS.
In the CMS, the final estimate for a count is $\min(h_i(\textit{key}))$ for each
row $i$.  The minimum makes sense here since collisions can only increase the
counts.
The Laplace distribution however is symmetric around zero, so we may add
negative noise to the counts.  Taking the minimum of multiple draws---each cell
is initiated with a random draw from the distribution---thus selects the most
extreme negative values, creating a downward bias that can be very large for a
small $\epsilon$.

We observe that because the mean of the Laplace distribution is 0, an unbiased
estimator would not suffer from this drawback. For tables with large noise, we
thus use a count-median sketch \cite{Charikar:2002:FFI:646255.684566}, which differs
in two ways: 1) each row $i$ has another hash function $s_i$
that maps the key to a random sign $s_i(\textit{key}) \in \{+1,-1\}$, with each cell
updated with $s_i(\textit{key})h_i(\textit{key})$; 2) the estimator is the median of all counts
multiplied by their sign, instead of the minimum.
The signed update means that collisions have an expected impact of zero, since
they have an equal chance of being negative or
positive, making the cell an unbiased estimate of the true count.
The median is a robust estimate that preserves the unbiased property.

Using this count-median sketch reduces the impact of noise, since values from the
Laplace distribution are exponentially concentrated around the mean of zero.
\S\ref{sec:evaluation} shows that for small $\epsilon$, or a large number of
features, it is worth trading the CMS's better guarantees for reduced noise impact
with the count-median sketch.

\subsubsection{Data Retention}
\label{sec:data-expiration}

While differential privacy provides a reasonable level of protection for past
observations, complete removal of information remains the cleanest, strongest
form of protection (design {\bf R3} in \S\ref{sec:requirements}).
\sysname supports data expiration with {\em windowed count tables}.
When an observation arrives, \sysname updates the count tables for the current
count window only.
To featurize $\x$, \sysname sums the relevant counts across windows.
Periodically, it drops the oldest window and invokes retraining of all models
in Velox ({\small \code retrain} method).
Our use of count-based featurization supports such behaviors because
retraining is cheap (\S\ref{sec:performance_eval}), so we can afford to do it
frequently.

\subsubsection{Count Selection}
\label{sec:count-table-selection}

\sysname seeks to support workload evolution (model changes/additions,
such as future model M4 in \F\ref{fig:architecture}) using only the widely
accessible stores without tapping into the historical raw data store.
To do so, it uses two approaches.
First, it stores the count tables in a very compact representation---the
count-median sketches---so it can afford to keep plenty of count tables.
Second, it includes an automatic process of {\em count table selection} that
inspects the data to identify {\em feature combinations} worth counting, 
whether they are used in the current workloads or not. 
This technique is useful because count featurization tends to obscure
correlations between features. For example, different users may have different
opinions about specific ads. Although that information could be inferred by
a learning algorithm from the raw data points, it is not accessible in the
count-featurized data unless
we explicitly count the joint occurrences of specific users with specific ads, i.e., maintain a table
for the $\langle userId, adId \rangle$ group.

We adapted several feature selection techniques \cite{guyon2003introduction} 
to select feature groups and describe one
here.
{\em Mutual Information} (MI) is a measure of dependence between two random variables.
A common feature selection technique keeps features of high MI with the label. 
We extend this mechanism for group count selection.
Our goal is to identify feature groups that provide more information
about the label than individual features.
For each feature $x_i$, we find all other features $x_j$ such that
$x_i$ and $x_j$ together exhibit higher MI with the label than $x_i$ alone.
From these groups, we select a configurable number with highest MIs.
To find promising groups of larger sizes, we apply this process greedily,
trying out new features with existing groups.
For each selected group, \sysname creates and maintains a count table.

This exploration of promising groups operates on the {\em hot window of raw
data}.
Because the hot raw data is limited, the selection may not be entirely
reliable.
Therefore, count tables for new groups are added on a ``trial basis.''
As more data accumulates in the counts, \sysname re-evaluates them by
computing the MI metric {\em on the count tables}.
With the increased amount of data, \sysname can make a more reliable decision
regarding which count tables to keep and which to drop.
Because count selection---like feature selection---is never
perfect, we give engineers an API to specify groups that they
know are worth counting from domain knowledge.
Finally, like the weight selection process, count selection should be made 
differentially private so the groups selected in a
particular hot window, which are preserved over time, do not leak information
about the window's data in the future.
\S\ref{s:dp-groups} proposes a method for making count selection private.

\subsection{Supported Workload Evolution}
\label{sec:workload-updates}

Count featurization is a model-independent preprocessing step, allowing \sysname to absorb some
common evolutions during an ML application's life cycle without tapping the historical raw data store.
\S\ref{sec:news_recommendation} gives anecdotal evidence of this claim from a production workload.
This section reviews the types of workload changes \sysname currently absorbs.

A developer may want to change four aspects of the model:
(1) the algorithm used to train the model (2) hyperparameters for the model or
for the underlying optimization algorithm, (3) features used by the model, and
(4) the predicted label.
\sysname supports (1) and (2), partially supports (3), and usually does not support (4).

\noindent
$\bullet$ {\em Algorithm changes: Supported.}
\sysname allows developers to move between types of models and libraries used to
train those models as long as they are using features and labels that are
already counted.
In our evaluation we experimented with linear models and neural networks in
Vowpal Wabbit~\cite{langford2007vowpal} and gradient boosted trees
in scikit-learn~\cite{scikit-learn} using the same count tables.

\noindent
$\bullet$ {\em Hyperparameter tuning: Supported.}
By far the most common type of model change we encountered, both in our own evaluation and in reports from a production setting, was hyperparameter tuning.
For example, a developer may want to change model hyperparameters, such as the
number of hidden units in a neural network, or tune parameters of the underlying optimization algorithm, such as the learning rate or an L1/L2 regularization penalty.
Changing hyperparameters is independent from the underlying features so is supported by \sysname.

\noindent
$\bullet$ {\em Feature changes: Partially supported.}
\sysname supports making minimal feature changes.
A developer may want to perform one of three types of feature changes: adding new features, removing existing features, or adding interactions between existing features.
\sysname trivially supports removing existing features, and lets developers add new features if they are based on existing ones.
For example, the developer could not create an $\langle Age,
Location \rangle$ feature interaction if the individual features were not
already counted together. Introducing new feature combinations
or interactions requires creating new count tables. This highlights the
importance of count selection to support workload evolution.

\noindent
$\bullet$ {\em Label changes: Mostly unsupported.}
Changes in predicted labels are not supported except if a new label is
a subset of an existing label.
For example, a news recommender could not start predicting retention time instead of clicks unless retention time was previously declared as a label.
As with features, \sysname can support label changes when the new label is a subset of an existing one. For example, if a label exists that tracks retention time in time buckets, \sysname can support new, coarser labels,
such as the three classes ``0 seconds,'' ``less than a minute,'' and ``more than a minute.''

\subsection{Summary}
\label{sec:design-summary}

With these components, \sysname
meets the design requirements noted in \S\ref{sec:requirements}, as follows.
{\bf R1:} By enhancing the training set with historical statistics gathered
over a longer period of time, we minimize the hot data.
{\bf R2:} By automatically identifying combinations of features worth
maintaining, we avoid having to access the historical raw data for
workloads that use the same observation streams to predict the same label.
{\bf R3:} By rolling the count windows and retraining the application models,
we support data retention policies, albeit at a coarse level.
\S\ref{sec:evaluation} evaluates {\bf R4}: accuracy and performance impact.

\section{Prototype}
\label{sec:implementation}

\sysname is implemented in 2600 lines of Scala, as a modular library. It
integrates into the feature engineering stage of an ML pipeline, before the
actual learning algorithms are invoked.
The modular backend allows count tables to be stored locally in memory or in a remote datastore such as Redis or Cassandra.

We integrated \sysname into the Velox model management system~\cite{velox} with
minimal effort, by adding/modifying around 500 lines of code.
The changes we made to Velox involve interposing on all of Velox's interfaces
that interact with raw data (e.g., adding observations, making predictions, and retraining).
Now prediction requests are passed through the \sysname featurization layer,
which performs count featurization.

One of Velox's key contributions is performing low latency predictions by
pushing models to application servers.
To enable low-latency predictions, \sysname periodically replicates snapshots of the central count tables to the application servers, allowing them to perform featurization locally.
\S\ref{sec:performance_eval} evaluates prediction performance in Velox/\sysname with and without this optimization.
\section{Evaluation}
\label{sec:evaluation}

We evaluate \sysname using different versions of three data-driven
applications: two ad targeting applications, two movie recommendation
applications, and MSN's production news personalization system. We
compare models on count-featurized data to state-of-the-art models trained on raw data,
and answer these questions:
\begin{enumerate}
\item[{\bf Q1.}] Can we accurately learn on less data using counts?
\item[{\bf Q2.}] How does past-data protection impact utility?
\item[{\bf Q3.}] Does counting feature groups improve accuracy? 
\item[{\bf Q4.}] How efficient is \sysname?
\item[{\bf Q5.}] To what problems does \sysname apply?
\end{enumerate}

Our evaluation yields four findings:
(1) On classification problems, count featurization lets models perform within
4\% of state-of-the-art models while training on less than {\em 1\% of the data}.
(2) Count featurization enables powerful nonlinear algorithms, such as neural
networks and boosted trees, that would be infeasible due to high-cardinality
features.
(3) Protecting individual past observations with differential privacy adds 1\% penalty to the accuracy, which remains within 5\% of state-of-the-art models. 
(4) \sysname's performance overheads are small.

\begin{table}
{
\scriptsize
\centering
\begin{tabular}{| p{2.5cm} | p{1.18cm} | p{0.48cm} | p{0.50cm} | p{1.6cm} |}
    \hline
    \bf{App} & \bf{Dataset} & \bf{Obs.} & \bf{Feat.} & \bf{Baseline} \\
    \hline
    {\bf Ad targeting (classification)} & Criteo Kaggle~\cite{criteoKaggle} & 45M & 39 & neural net in Kaggle \cite{criteoKaggleWinner} \\
    \hline
    {\bf Ad targeting (classification)} & Criteo Full~\cite{criteoFull} & 1.2B & 39 & regularized linear model \\
    \hline
    {\bf Movie recommendation (classification)} & MovieLens \cite{movielens} &
    22M & 21 & matrix factorization~\cite{langford2007vowpal} \\
    \hline
    {\bf Movie recommendation (regression)} & MovieLens \cite{movielens} & 22M &
    21 & matrix factorization~\cite{langford2007vowpal}\\
    \hline
    {\bf News personalization (regression)} & MSN.com production & 24M & 507 &
    contextual bandits~\cite{Langford-www10,DR11}
    \\
    \hline
\end{tabular}
}
\vspace{-0.15cm}
\caption{\footnotesize {\bf Workloads.}
Apps and datasets; number of observations and features in each
dataset; and baselines used for comparison.
All baselines are trained using VW~\cite{langford2007vowpal}.
}
\label{t:datasets}
\vspace{-0.3cm}
\end{table}

\subsection{Methodology}
\label{s:methodology}
\begin{table}
{
\scriptsize
\centering
\begin{tabular}{|p{1.5cm}|p{2.0cm}|p{3.4cm}|}
    \hline
    \bf{Dataset} & \bf{Model} & \bf{Parameters} \\
    \hline
    \multirow{3}{*}{\bf{Criteo-Kaggle}}
    & {\bf B:} neural net (nn)& VW. One 35 nodes hidden layer with tanh activation. LR: 0.15. BP: 25. Passes: 20. Early Terminate: 1.\\
            \cline{2-3}
            & logistic regression (log. reg.) & VW. LR: 0.5. BP: 26.\\
            \cline{2-3}
            & gradient boosted trees (gbt) & Sklearn. 100 trees with 8 leaves. Subsample: 0.5. LR: 0.1. BP: 8.\\
   \hline
   \bf{Criteo-Full} & {\bf B:} ridge regression (rdg. reg.) & VW. L2 penalty: $1.5e^{-8}$. LR: 0.5. BP: 26.\\
    \hline
    \multirow{3}{*}{\pbox{1.5cm}{\bf{MovieLens Regression}}}
    & {\bf B: } singular value decomposition (svd) & VW. Rank 10. L2 penalty: 0.001. LR: 0.015. BP: 18. Passes: 20. LR Decay: 0.97. PowerT: 0.\\
            \cline{2-3}
            & linear regression (lin. reg.) & VW. LR: 0.5. BP: 22. Passes: 5. Early Terminate: 1.\\
            \cline{2-3}
            & gradient boosted trees (gbt) &  Sklearn. 100 trees with 8 leaves. Subsample: 0.5. LR: 0.1. BP: 8.\\
    \hline
    \multirow{3}{*}{\pbox{1.5cm}{\bf{MovieLens Classification}}}
    & {\bf B: } singular value decomposition (svd) & VW. Rank 10. L2 penalty: 0.001. LR: 0.015. BP: 18. Passes: 20. LR decay: 0.97. PowerT: 0.\\
            \cline{2-3}
            & logistic regression (log. reg.) & VW. LR: 0.5. BP: 22. Passes: 5. Early Terminate: 1.\\
            \cline{2-3}
            & gradient boosted trees (gbt) &  Sklearn. 100 trees with 8 leaves. Subsample: 0.5. LR: 0.1. BP: 8.\\
    \hline
   \bf{MSN.com} & contextual bandit & VW. IPS context. bandit. LR: 0.02. BP: 18.\\
    \hline
\end{tabular}
}
\vspace{-0.15cm}
\caption{\footnotesize {\bf Model parameters.}
The libraries and parameters used to train each model.
The parameters not noted use library defaults.
``LR'' indicates the learning rate.
``BP'' indicates the hash featurization's bit precision (only applicable to raw models).
``PowerT'' exponent controls learning learning rate decay per step.
``{\bf B:}'' indicates that the model will be used as a baseline.
VW and Sklearn denote that the model was trained with Vowpal Wabbit~\cite{langford2007vowpal} and scikit-learn~\cite{scikit-learn}, respectively.
}
\label{t:models}
\vspace{-0.3cm}
\end{table}

\headingg{Workloads.}
Table~\ref{t:datasets} shows our apps, datasets, and baselines. We
defer discussion of MSN to \S\ref{sec:news_recommendation}.

\noindent
$\bullet$
{\em Criteo ad targeting.}
Using two versions of the well-known Criteo ads dataset, we build a binary
click/no-click classifier. We use seven days of the Criteo ad click dataset
amounting to 1.2 billion total observations.
This dataset is very imbalanced with an approximate click rate of 3.34\%.
The second version of the Criteo dataset has 45 million observations, and was released as part of a Kaggle competition.
In the Criteo Kaggle dataset, the click and non-click points were sampled at
different rates to create a more balanced class split with a 25\% click rate.
Each observation has 39 features (13 numeric, 26 categorical), and 8 of the
categorical features are high dimensional ($>100K$ values).
The numeric features were binned into 4 equal size bins for each dataset.
As a baseline, we use a feed-forward neural network that performed well for the
competition dataset~\cite{criteoKaggleWinner}, and we use ridge regression for
the full dataset.

\noindent
$\bullet$
{\em MovieLens movie recommendation.} Using the well-known MovieLens dataset,
which consists of $22M$ ratings on $34K$ movies from $240K$ users, we build
two predictors: (1) a regression model that predicts the user's rating as a continuous value in $[0,5]$,
(2) a binary classifier that predicts if a user will give a rating of 4 or more.
As a baseline, we use the matrix factorization algorithm in Vowpal Wabbit
(VW)~\cite{langford2007vowpal}; algorithms in this class are state-of-the-art
for recommender systems~\cite{koren2009matrix}, although this specific implementation is not the most advanced.

\heading{Method.}
For each application, we try a variety of count models, including linear
or logistic regression, neural networks, and boosted trees.
We split each dataset by time into a training set (80\%) and testing set (20\%),
except for the full Criteo dataset for which
we use the first six days for training and the seventh for testing.
On the training set, we compute the counts and train our models on windows of
growing sizes, where all windows contain the most recent training data and grow backwards to include older data.
This ensures that training occurs on the most recent data (closest to the
testing set), and that count tables only include observations from the hot
window or the past.
We use the testing set to compare the performance of our count algorithms to their raw data counterparts and to the baseline algorithms.
For all baselines, we apply any dimensionality reduction mechanisms (e.g., hash
featurization~\cite{weinberger2009feature}) that those models typically apply to
strengthen them.

\begin{figure*}[t!]
\subfigure[{\bf MovieLens classification}]{
  \includegraphics[width=0.31\linewidth]{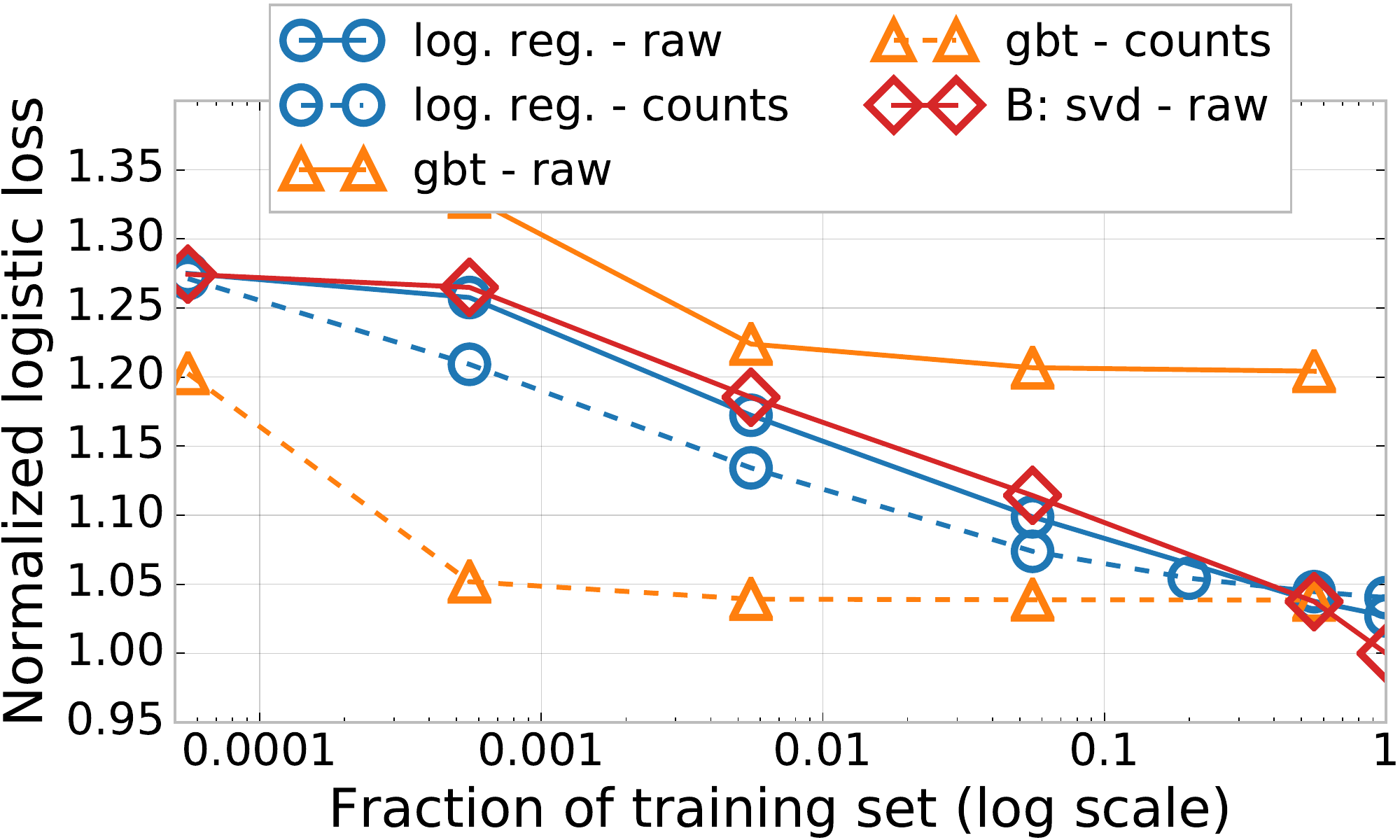}
  \label{f:counts_raw_movielens}
}
\subfigure[{\bf Criteo-Kaggle classification}]{
  \includegraphics[width=0.31\linewidth]{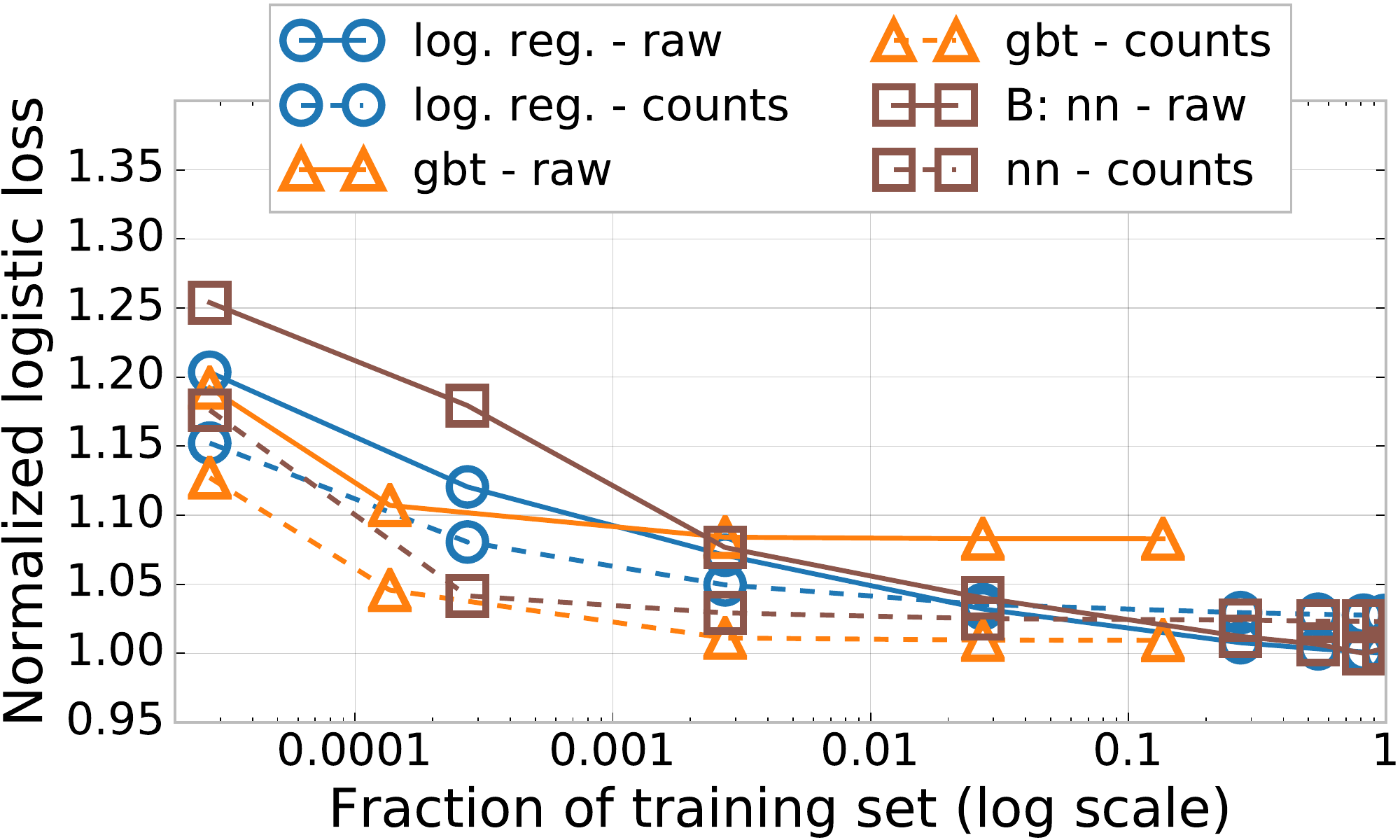}
  \label{f:counts_raw_criteo}
}
\subfigure[{\bf Criteo-Full classification}]{
  \includegraphics[width=0.31\linewidth]{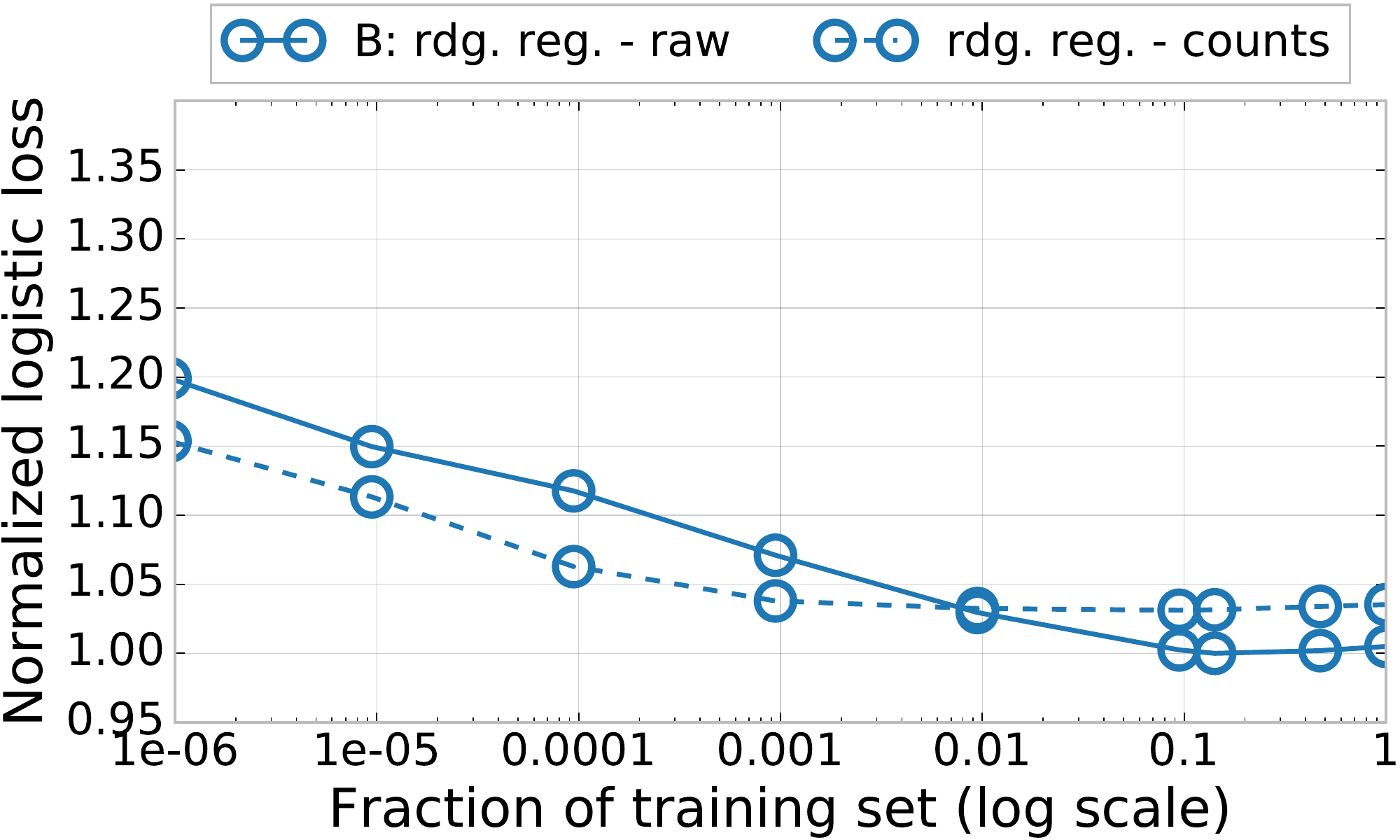}
  \label{f:counts_raw_criteo_full}
}
\vspace{-0.35cm}
  \caption{\footnotesize {\bf Normalized losses for raw and count algorithms.}
    ``B:'' denotes the baseline model.
    Count algorithms converge faster than raw data algorithms, to results that are within 4\% on MovieLens, and within 2\% and 4\% on Criteo Kaggle and full respectively.
  }
\label{f:counts_vs_raw}
\vspace{-0.3cm}
\end{figure*}

\heading{Metrics.}
We use two model accuracy metrics.

\noindent 
(1) The {\em average logistic loss} for classification problems with categorical labels (e.g. click/no-click).
Algorithms predict a probability
for each class and are penalized by the logarithm of the
probability predicted for the true class: $-\log(p_{\textit{true\_class}})$.
Models are penalized less for incorrect, low-confidence predictions and more for
incorrect, high-confidence predictions.
Logistic loss is better suited than accuracy for classification problems with imbalanced classes because a model cannot perform well simply by returning the most common class.

\noindent 
(2) The {\em average squared loss} for regression problems with continuous labels. Algorithms make real-valued predictions that are penalized by the square of the difference with the label: $||\textit{prediction} - \textit{label}||^2$.

We conclude our evaluation with our experience with a production setting, in which we can directly estimate click-through rate, a more intuitive metric.

\heading{Result interpretation.}
All graphs report loss normalized by the baseline model trained on the {\em entire training data}.
Lower values are better in all graphs: a value of $1$ or less means that we beat the baseline's best performance; and a value $>1$ means that we do worse than the baseline.

For completeness, we specify our baselines' performance:
MovieLens classification matrix factorization has a logistic loss of 0.537;
MovieLens regression matrix factorization has a squared loss of 0.697;
Criteo-Kaggle neural network has a logistic loss of 0.467;
and Criteo-Full ridge regression has a logistic loss of 0.136.

\subsection{Training Set Reduction (Q1)}

\begin{figure*}[t!]
\subfigure[{\bf MovieLens boosted tree}]{
  \includegraphics[width=0.31\linewidth]{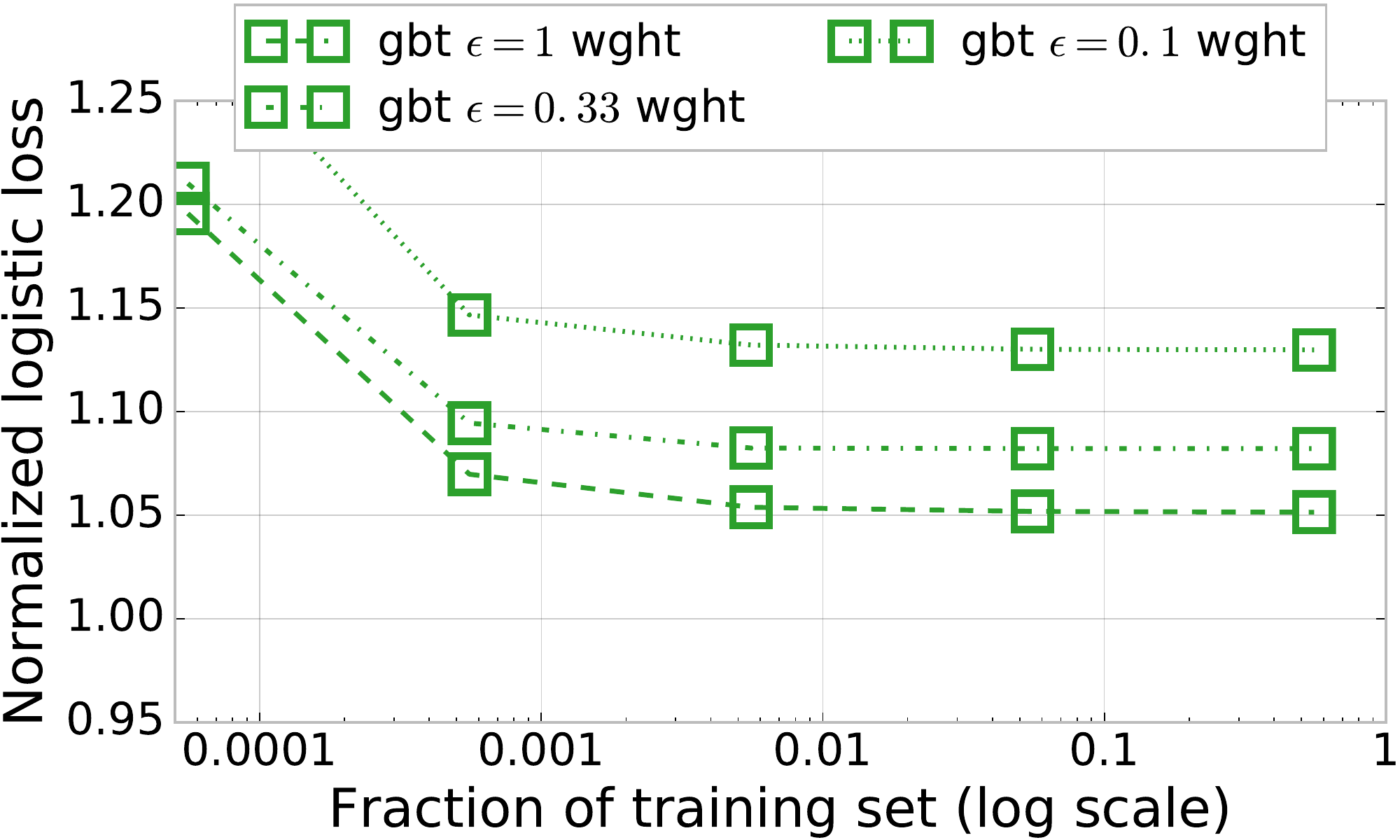}
  \label{f:noise_movielens_tree}
}
\subfigure[{\bf Criteo-Kaggle algorithms}]{
  \includegraphics[width=0.31\textwidth]{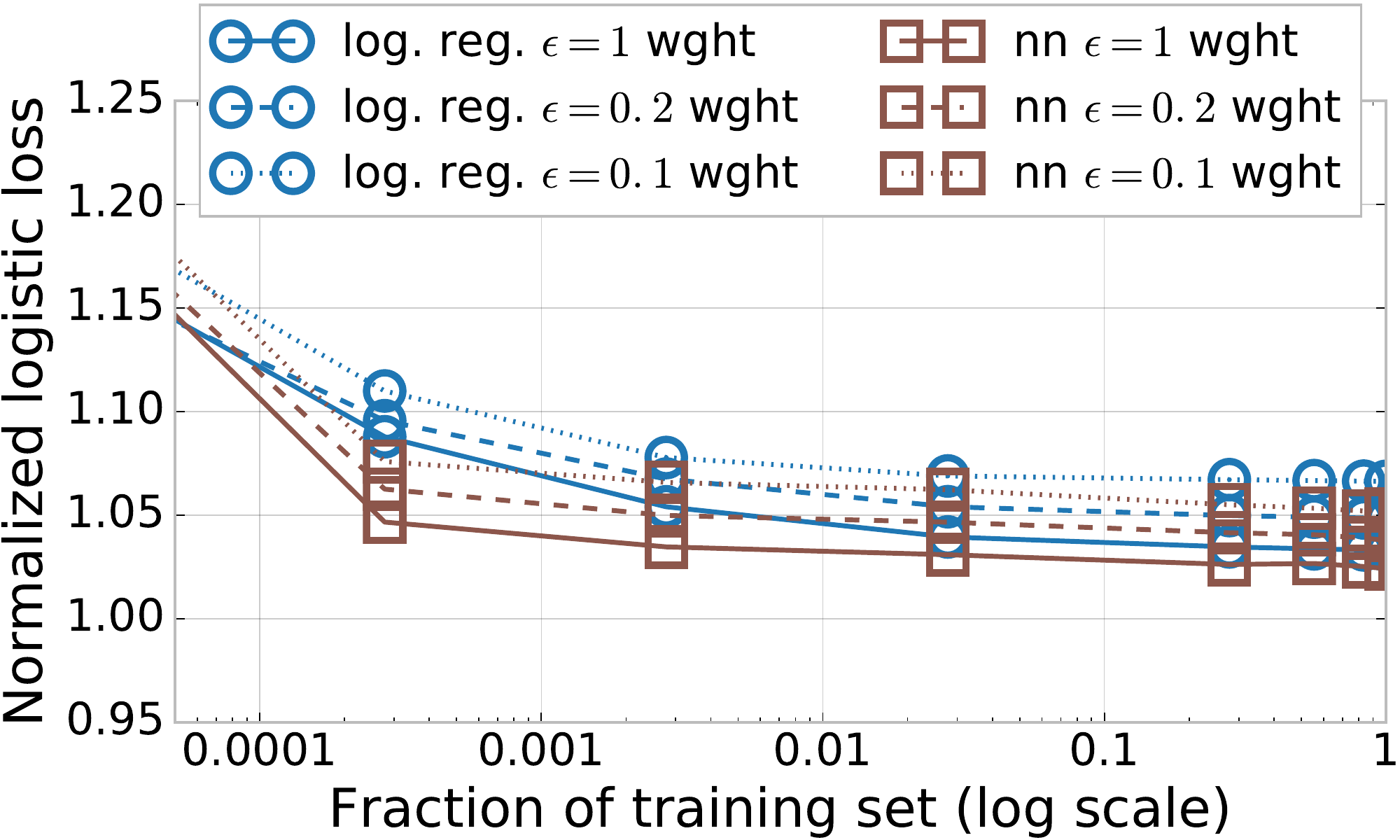}
  \label{f:noise_criteo_kaggle_nn}
}
\subfigure[{\bf Criteo-Full ridge regression}]{
  \includegraphics[width=0.31\textwidth]{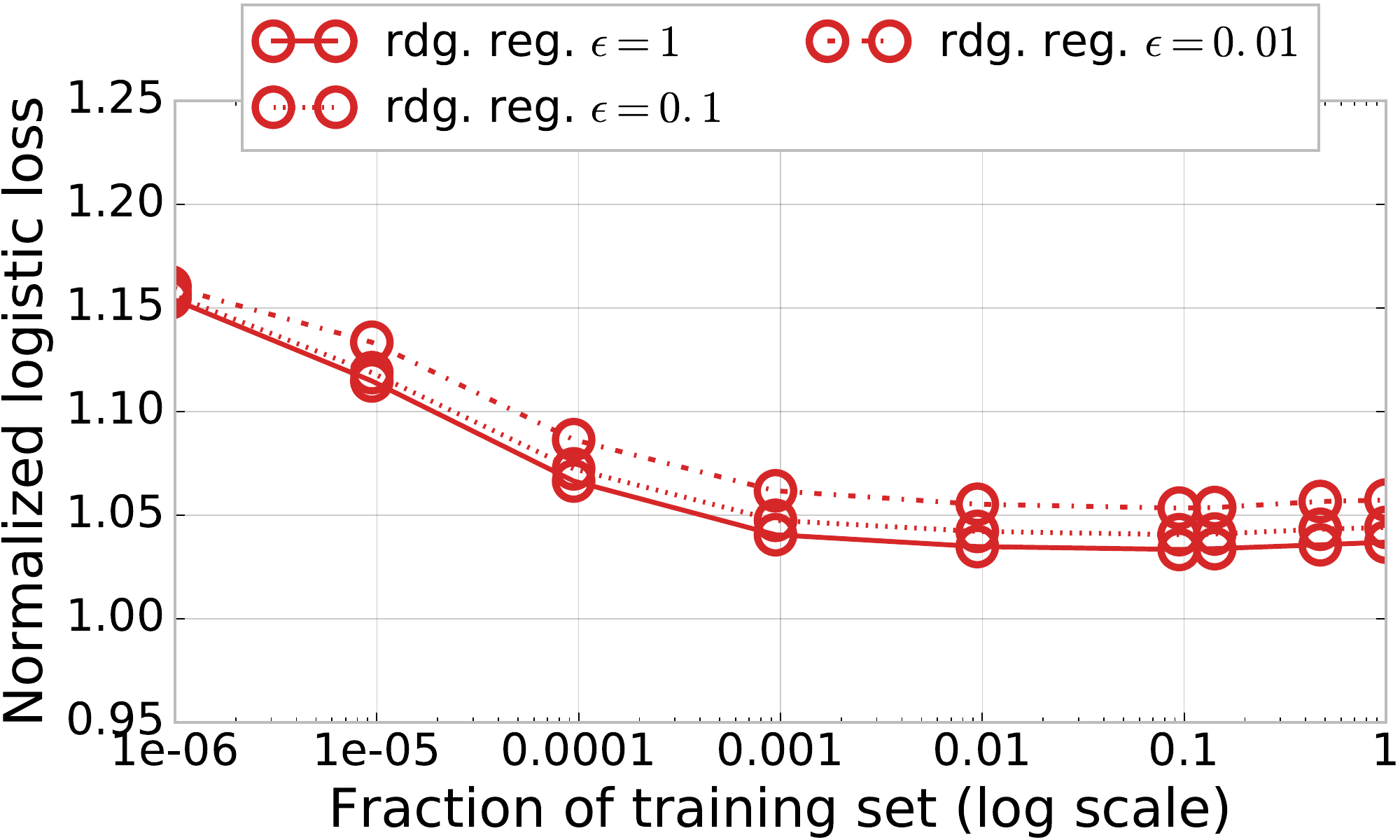}
  \label{f:noise_criteo_full}
}
\vspace{-0.35cm}
\caption{\footnotesize {\bf Impact of data protection.}
  Results are normalized by the baselines.
  We fix $k=1$ and vary $\epsilon$, the privacy budget.
  \F\ref{f:noise_movielens_tree} and \F\ref{f:noise_criteo_kaggle_nn} show results using the weighted noise (denoted wght).
  On MovieLens our weighting scheme is crucial to hide $1$ observation.
  On Criteo we can easily hide $1$ observation with little performance degradation and can hide up to $100$ observations while remaining within 5\% of the baseline.
}
\label{f:noise_impact}
\vspace{-0.5cm}
\end{figure*}

\sysname's design is predicated on count featurization's ability to substantially reduce training sets.
While this method has long been known, we are unaware of scientific studies of its effectiveness for training set reduction.
We hence perform a study here.
The count models must converge faster than raw-data models
(reach their best performance with less data), and perform on par with state-of-the-art baselines.
\F\ref{f:counts_vs_raw} shows the performance of several linear and nonlinear
models, on raw and count-featurized data. We make two observations.

First, {\bf training with counts requires less data}.
On both Criteo and MovieLens the best count-featurized algorithm approaches
the best raw-data algorithm by training on {\em 1\% of the data or
less}.
On Criteo-Kaggle (\F\ref{f:counts_raw_criteo}), the count-featurized neural network comes within 3\% of the baseline when trained on 0.4\% of the data and performs within 1.7\% of the baseline with 28\% of the training data.
On Criteo-Full (\F\ref{f:counts_raw_criteo_full}), the count-featurized ridge
regression model comes within 3.3\% of the baseline with only 0.1\% of the data, and within 2.5\% when trained on 15\% of the data.
These results show that models trained on count-featurized data can perform
close to raw models in both balanced and very imbalanced datasets (Criteo Full
and Kaggle's respective click rates are 3\% and 25\%).
On MovieLens (\F\ref{f:counts_raw_movielens}), the count-featurized boosted tree needs only
0.8\% of the data to get within 4\% of the baseline, or match the raw data logistic regression.
Because counts summarize history and reduce dimensionality, they allow algorithms to perform well with very little data. We say that they {\em converge faster} than raw data algorithms.

Second, \textbf{counts enable new models}.
In \F\ref{f:counts_vs_raw}, the boosted tree performs poorly on
raw data but very well on the count-featurized data.
This reveals an interesting insight.
The raw-data boosted tree uses a dimensionality reduction technique known as feature hashing~\cite{weinberger2009feature}, which hashes all categorical values to a limited-size space.
This technique exhibits a trade-off: increasing the hash space reduces
collisions at the cost of introducing more features, leading to overfitting.
Count featurization does not have this problem: a categorical feature is mapped to a few new features (roughly one per label value).
This lets us train boosted trees very effectively.

\subsection{Past-Data Protection Evaluation (Q2)}
\label{s:noise_eval}

We have shown that count-featurized algorithms converge faster than models trained on raw data. This allows \sysname to keep, and thus expose, only a small amount of raw data to train ML models.
However the count tables, while only aggregates of past data, can still leak information about past observations.
To prevent such leaks, \sysname adds differentially private noise to the tables.
The amount of noise to add depends on the desired privacy guarantee, parameterized by $\epsilon$ (smaller is more private), but also on the  number of features (see Table~\ref{t:datasets}) and CMS hash functions (five here), through the formula from \S\ref{sec:differential-privacy}.
In this section we evaluate the noise's impact on performance, as well as
\sysname's two mechanisms that increase data utility: automatic weighted noise
infusion and the use of private count-median sketches.
We also show the impact of the number of windows used, which
defines the granularity at which past observations can be entirely dropped.

\noindent
\textbf{Impact of noise.}
\F\ref{f:noise_impact} shows the performance of different algorithms and
datasets when protecting an observation, $k=1$, with different privacy budgets $\epsilon$ (note the direct tradeoff between the two parameters: the noise is proportional to $\frac{k}{\epsilon}$).
We find that \sysname can protect observations with minimal performance loss.
When $\epsilon=1$, the boosted tree model on the MovieLens dataset remains
within 5\% of the baseline with only 1\% of the training data.
The logistic regression and neural network models on the Criteo-Kaggle dataset perform within 2.7\% and 1.8\% of the baseline respectively, and the Criteo-Full ridge regression is within 3\%.
All Criteo models also come within 5\% of their respective baseline with a
privacy budget as small as $\epsilon=0.2$.

The Criteo-Full ridge regression performance degrades less than models on other datasets when the noise increases.
For instance, it degrades by less than 1\% with $\epsilon$ going from 1 to 0.1, while the Criteo-Kaggle neural network loses 6.5\%.
This is explained by the fact that the amount of noise required to make a query differentially private is not related to the size of the dataset.
The Criteo-Full dataset is much larger, so the additional noise is much smaller relative to the counts.

\begin{figure*}[t!]
\subfigure[{\bf MovieLens boosted tree}]{
  \includegraphics[width=0.31\linewidth]{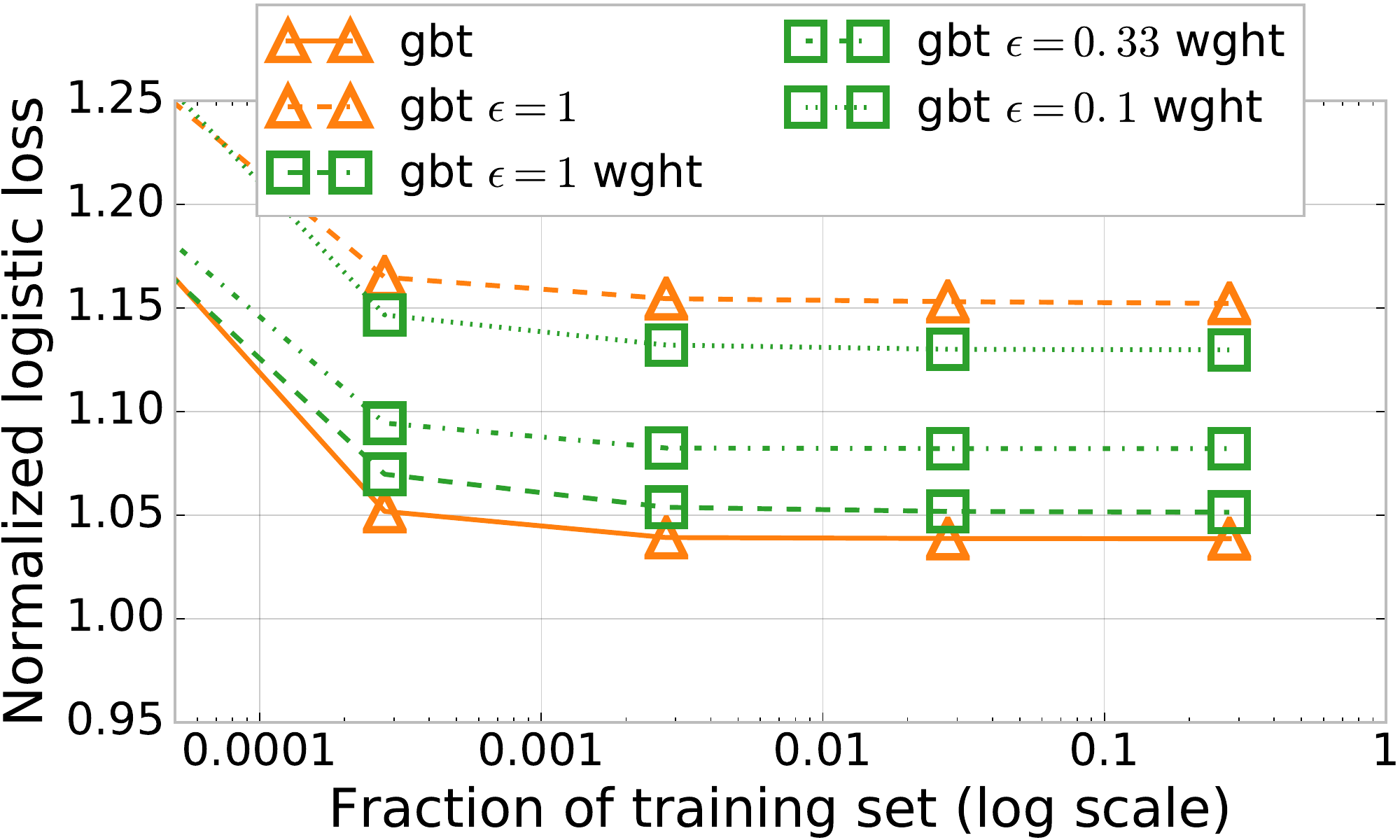}
  \label{f:movielens_noise_weighted}
}
\subfigure[{\bf Criteo-Kaggle neural network}]{
  \includegraphics[width=0.31\textwidth]{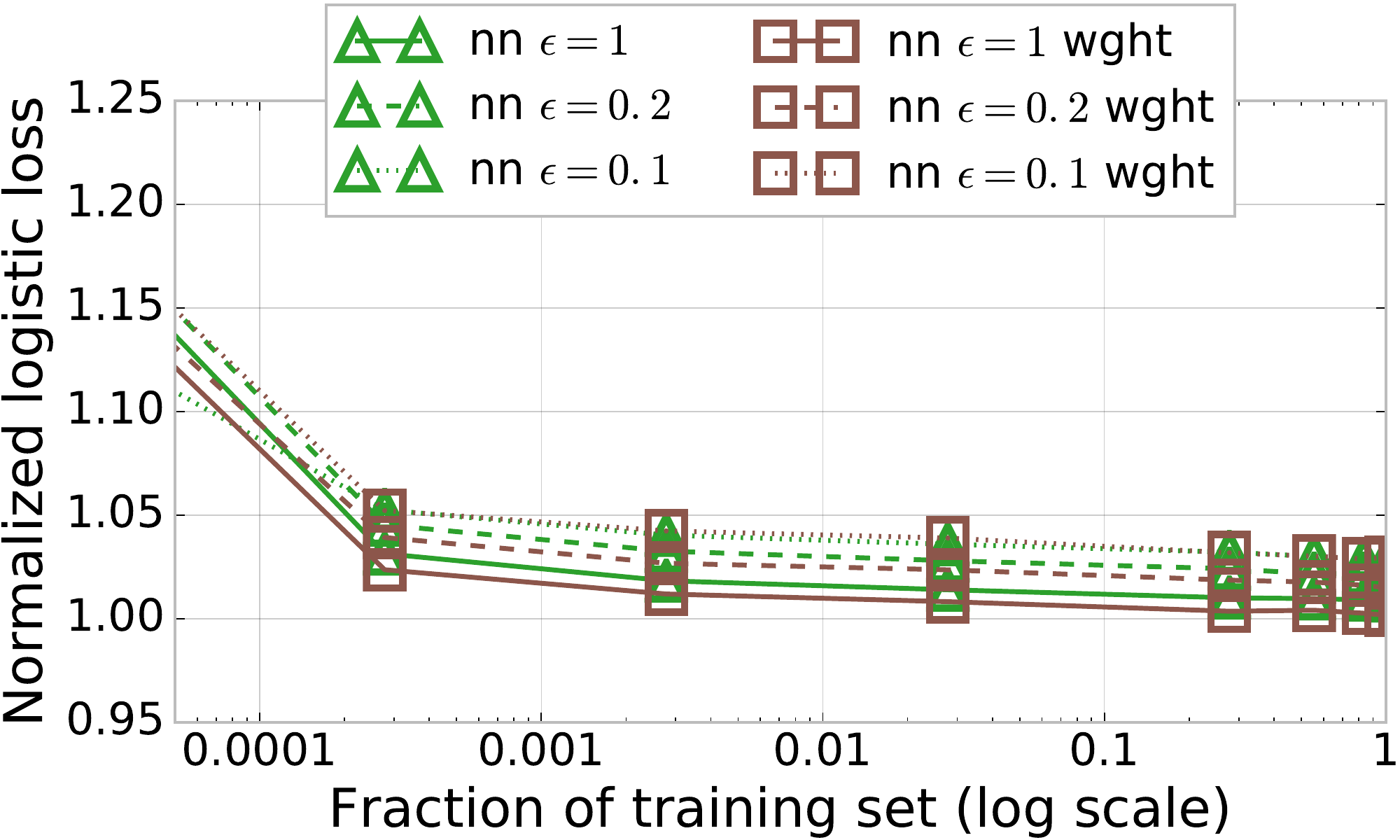}
  \label{f:crite_kaggle_noise_weighted}
}
\subfigure[{\bf Sketch comparison}]{
  \includegraphics[width=0.31\linewidth]{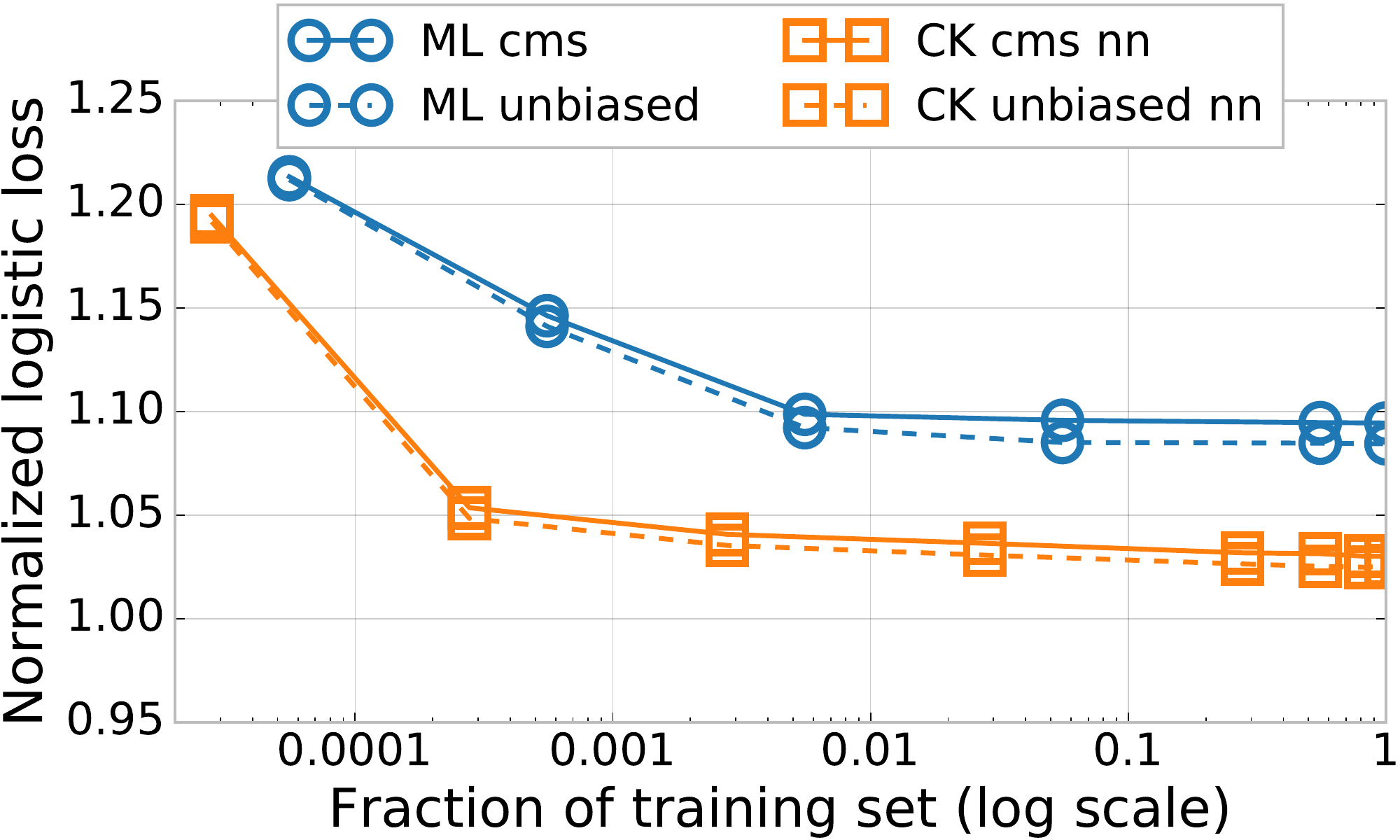}
  \label{f:datastructure_performance}
}
\vspace{-0.35cm}
\caption{\footnotesize {\bf Impact of data protection (continued).}
    Results are normalized to the baselines.  We fix $k=1$ and vary $\epsilon$, the privacy budget.
    (a) Without the feature weighting trick the gradient boosted trees perform unacceptably poorly.
    (b) The weighting trick marginally improves the performance of Criteo-Kaggle models over equally distributing the privacy budget.
    (c) Private count-median sketch improves performance in both MovieLens (ML)
    and Criteo-Kaggle (CK) models with $\epsilon=1$.
}
\label{f:noise_mitigation}
\vspace{-0.45cm}
\end{figure*}

\noindent
\textbf{Weighted noise infusion.}
Weighted noise infusion is integral to the protection of past observations with minimal performance cost.
\F\ref{f:movielens_noise_weighted} shows the impact of noise on the boosted
tree for the MovieLens dataset. Without weighting the privacy budget of different features, the model performs 15\% worse than the baseline even for $\epsilon=1$.
With non-private weighting, the MovieLens model performs at 5\% of the baseline.
The weighted noise infusion technique is thus critical to maintaining performance on the MovieLens dataset.
Intuitively, this is because the users making the rating and the movie being
rated are the most important features when predicting ratings.
Most users rate relatively few movies, and a long tail of movies are rarely rated, so their respective counts are quickly overwhelmed by the noise when the privacy budget is equally distributed among all features.

The Criteo models do not depend as much on the weighting trick, since they do not rely
on a few features with small counts.
Noise weighting is still beneficial, though: e.g., the Criteo-Kaggle neural
network gains about 0.5\% of performance, as shown in
\F\ref{f:crite_kaggle_noise_weighted}.

\begin{figure*}[t]
  \begin{minipage}{0.30\textwidth}
  \centering
  \footnotesize
    \includegraphics[width=\textwidth]{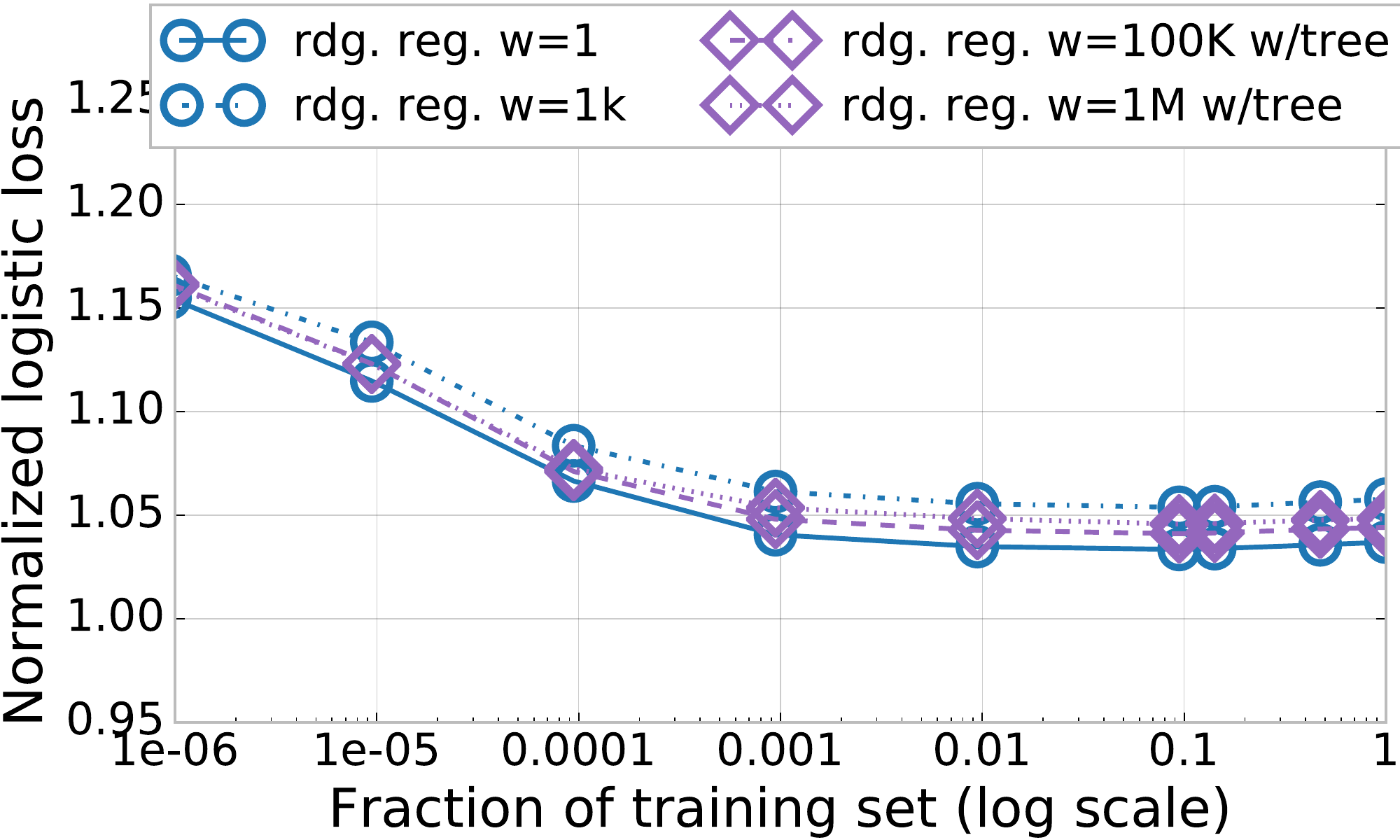}
    \caption{{\footnotesize{\bf Criteo-Full windows.} The Criteo datasets can
    support 1K windows with reasonable penalty. Supporting more windows requires a scheme based on binary trees.}}
    \label{f:criteo_full_noise_tree}
  \end{minipage}
  \hspace{0.15cm}
  \begin{minipage}{0.30\textwidth}
    \centering
    \footnotesize
    \includegraphics[width=\linewidth]{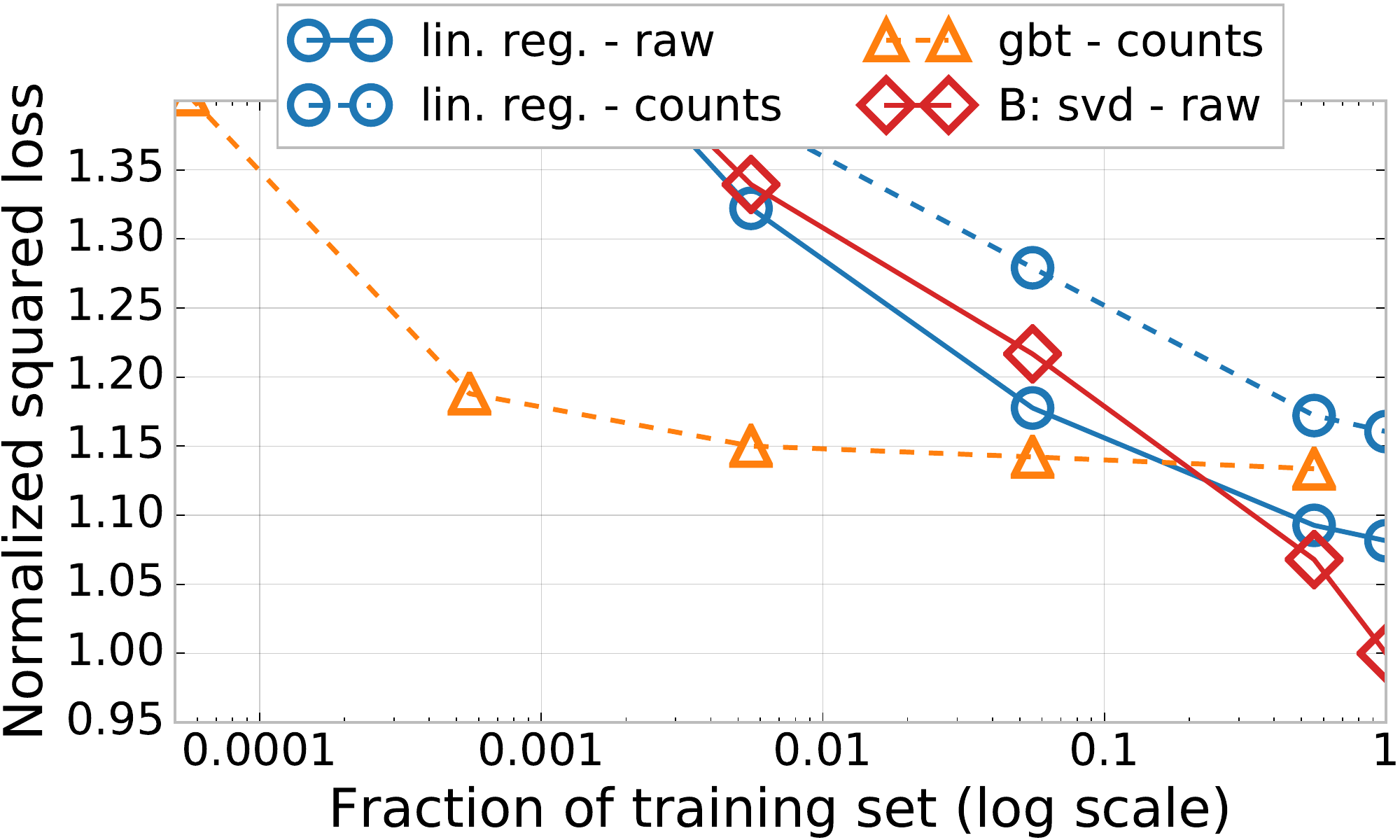}
    \caption{{\footnotesize{\bf MovieLens regression.}
    Linear regression algorithms are not amenable. Boosted tree converges
    quickly but does not match the baseline.}}
    \label{f:regression_movielens}
  \end{minipage}
  \hspace{0.15cm}
  \vspace{0.5cm}
  \begin{minipage}{.35\textwidth}
    \centering
    \vspace{0.4cm}
    \setlength\tabcolsep{0.05cm}
    {\scriptsize
    \begin{tabular}[d]{|c|c|c|c|}
      \hline
      Action & P. w/o cache & P. w/ cache & Velox \\
      \hline
      Featurization & 99.22\% & 4.37\% & N/A \\
      \hline
      Marshalling & 0.04\% & 6.44\% & 7.06\% \\
      \hline
      Prediction & 0.01\% & 0.51\% & 0.63\% \\
      \hline
      Network/Framework & 0.73\% & 88.68\% & 92.31\% \\
      \hline
      Total Latency & 283.69 ms & 1.65 ms & 1.58 ms \\
      \hline
    \end{tabular}
    }
    \vspace{0.45cm}
    \caption{{\footnotesize {\bf Prediction Latency.} Median time to serve a model prediction. Caching is crucial for \sysname to achieve low overhead compared to Velox.}}
    \label{t:performance_comparison}
  \end{minipage}%
\vspace{-0.8cm}
\end{figure*}

\noindent
\textbf{Private count-median sketch.}
Another technique that \sysname uses to reduce the impact of noise is to switch
to a private count-median sketch.
As noted in~\S\ref{sec:differential-privacy}, the count-min sketch will exhibit a strong downward bias when initialized with differentially private noise, because taking the minimum of multiple observations will select the most extreme negative noise values.
The count-median sketch uses the median instead of the minimum and does not suffer from this effect.
\F\ref{f:datastructure_performance} shows that when noise is added, the count-median sketch improves performance over the count-min sketch by around 0.5\%, on MovieLens and Criteo-Kaggle.

When combined with weighted noise infusion, the private count-median sketch is less
useful at first, since the noise is small on features with small counts.
However, it provides an improvement for lower $\epsilon$. For
instance, the MovieLens boosted tree improves by 0.5\% even after noise
weighting for $\epsilon=0.10$.

\noindent
\textbf{Number of windows.}
Another factor impacting accuracy is the number of count windows kept to
support granular retention policies.
\F~\ref{f:criteo_full_noise_tree} shows Criteo-Full's ridge regression for $k=1$ and $\epsilon=1$ while varying the number of windows.
We observe that it is possible to support a large number of windows.
On Criteo, we can support $1000$ windows with little degradation, enough to support a daily granularity for a multi-year retention period.
While we believe this granularity for retention policies should be enough in
practice, we also simulated a binary tree scheme
\cite{Chan:2011:PCR:2043621.2043626} that supports huge numbers of windows. We
can see that on Criteo, this allows using $100K$ windows with a penalty similar
to $10$ windows using the basic scheme.

\subsection{Count Selection Evaluation (Q3)}
\label{s:count_selecton_evaluation}

\noindent
\textbf{Without noise.}
We measure the performance of our algorithms when the featurization is
augmented by MI-selected groups.
We evaluate on MovieLens, as groups provided little additional benefit on
Criteo. A total of 35 groups were selected by MI and given 10\% of
the privacy budget to share. 
When using these groups, the accuracy of the count
boosted tree gets within 3\% of the baseline with the same 0.8\% of the data, 1\% better than without feature groups.
Logistic regression does not improve asymptotically but converges faster,
getting within 5\% of the baseline with 15\% of the data instead of 22\%.
Thus, count selection selects relevant groups.

\noindent
\textbf{With noise.}
We also evaluate the impact of group selection on MovieLens with noise
$k=1$, $\epsilon =1$.
Logistic regression is not improved by the grouped features, but the boosted
tree is still 1\% closer to the baseline.
Thus, the algorithm can still extract useful information from the groups despite
the increased noise.

While these results are encouraging, we leave for future work the full investigation
of how the improvement in accuracy gained from maintaining and using relevant groups
is affected by the higher noise levels necessary to maintain a large number of count tables
for fixed $\epsilon$.

\subsection{Performance Evaluation (Q4)}
\label{sec:performance_eval}

We evaluate \sysname's overhead on Velox by measuring the median
latency of a prediction request to Velox.
We perform this evaluation using the 39-feature Criteo dataset.
\F\ref{t:performance_comparison} shows the median latencies and a
breakdown of the time into four components: computing the prediction,
unmarshalling the message into a usable form, performing count featurization, and
other functions like the network and traversing the web stack.
We show the results with and without count table caching in the application
servers (\S\ref{sec:implementation}).
Without caching, prediction latency is around 200ms.
Caching reduces it to 1.6ms, a 5\% overhead with the total time dominated by the network and traversing the web framework used to implement Velox.
Pushing count tables to the application servers is crucial for
performance and does not significantly increase the attack surface.

\subsection{Applicability Evaluation (Q5)}
\label{sec:applicability_evaluation}

\sysname works well for classification problems.
We now consider another broad class of supervised learning problems:
regression problems. In regression, the algorithm guesses a label on a
continuous scale, and the goal is for the prediction to be as close to the true label as possible.
Intuitively, count featurization should be less effective for regression
problems, because it needs to bin the continuous label
into discrete buckets.

Fig.~\ref{f:regression_movielens} shows the performance of linear and boosted tree (nonlinear) regressions on the MovieLens dataset.
We first observe that linear regression does worse on count-featurized data than
on raw data. This is not surprising: count featurization gives the probability
of each label conditioned on a feature. The algorithm cannot find a linear
relationship between, say, $P(\textit{rating}=3 | \textit{user})$ and the rating.
Indeed, the rating does not keep growing with this probability, it
keeps getting closer to 3.  

Nonlinear algorithms do not have this limitation. The boosted tree
converges quickly and outperforms raw models trained on similar amounts of data
until we reach 55\% of the data. At that point, the boosted tree plateaus and
never comes close to the baseline.  Although we did not find good algorithms for
this dataset, we suspect that some nonlinear algorithms may perform well on counts.

Count featurization is most reminiscent of the counts used by Naive Bayes
classifiers~\cite{russel-norvig-ai}, and there are workloads for which it is not suitable.
For instance, count featurization requires a label and is thus not applicable to
unsupervised learning. Other feature representations may be better suited to
such types of models. Our choice of count featurization reflects its suitability
to data protection in a practical system architecture.

Even in settings that are less amenable to \sysname, such as online learning
applications that avoid retraining, we found that \sysname can perform well and
help protect past observations, as we describe in the next section.

\subsection{Experience with a Production Setting}
\label{sec:news_recommendation}

\begin{figure}[t]
  \centering
  \footnotesize
    \includegraphics[width=0.4\textwidth]{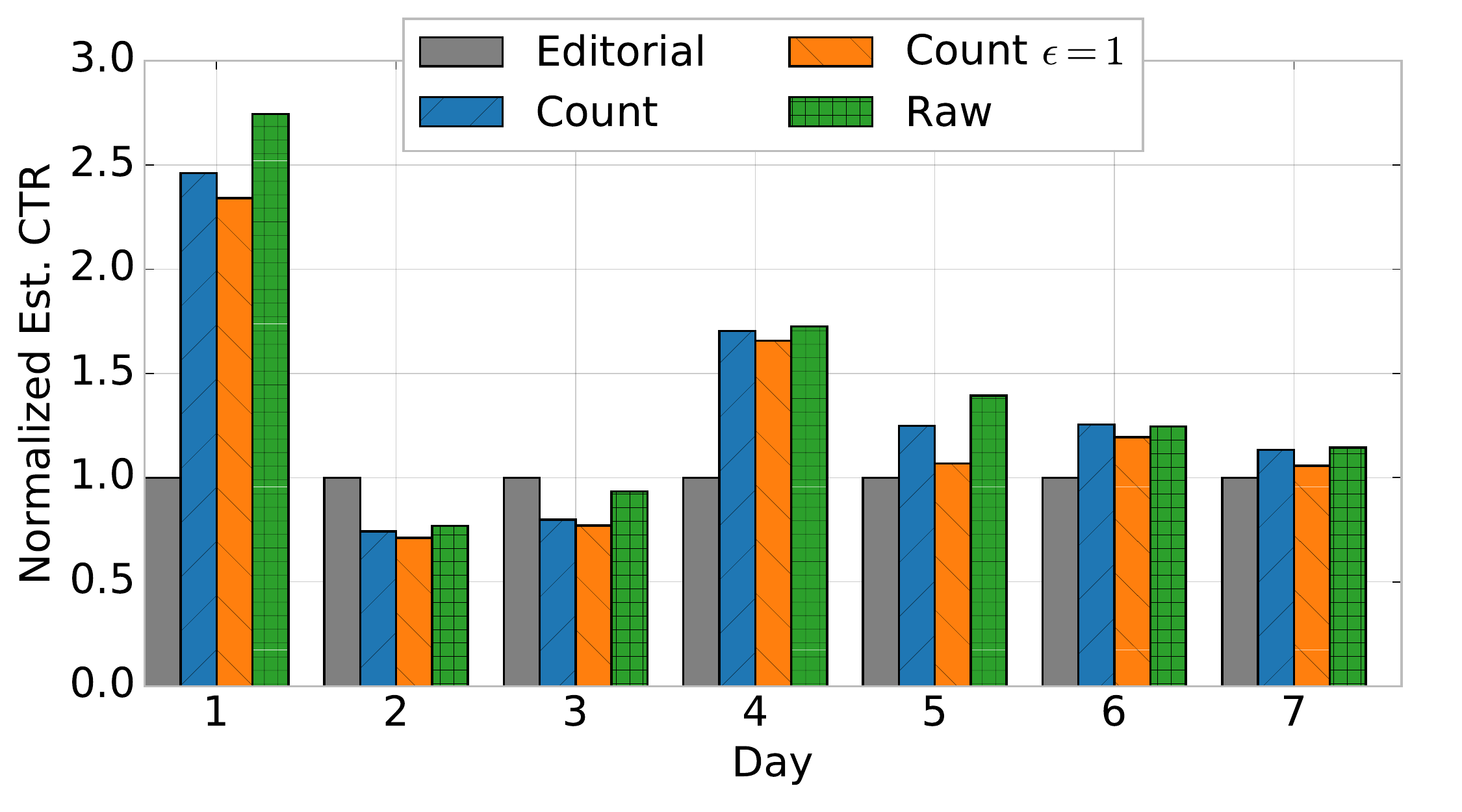}
\vspace{-0.3cm}
     \caption{\footnotesize {\bf Estimated article CTR for MSN.} The raw model,
     count model, and private count model are normalized against the estimated
     performance of human editors. The count models perform slightly worse than
     the raw models; all models outperform human editors on five out of
     seven days.}
    \label{f:news_progressive_loss}
\vspace{-0.5cm}
\end{figure}

In addition to public datasets, we also evaluated \sysname on
a production workload.  One of the authors helped build MSN's news personalization service, which we used to evaluate three aspects: 
(1) How to adapt count featurization to a different type of
learning, (2) how \sysname applies to this application, and 
(3) how \sysname supports the application's workload evolution.

\heading{Adapting count featurization.}
MSN uses contextual bandit
learning~\cite{Langford-nips07,monster-icml14} 
(via the Decision Service~\cite{mwtDecisionService2016}) 
to personalize the order of articles shown to each user so as to maximize
clicks, based on 507 features of user demographics and past browsing history.  
This is a challenging scenario due to
the large number of features and low click signal.
Contextual bandit algorithms
use randomization to explore different action choices, e.g., picking the top
article at random. This produces a dataset that assigns a probability
(importance weight) to each datapoint. The probabilities are used to optimize
models offline and obtain unbiased estimates of their performance had they been
run online~\cite{DR11,Langford-www10,Langford-wsdm11}.

Importance-weighted data have interesting implications for Pyramid.
When updating the count tables with a given data point, \sysname
must increment the counts by $1/p$, rather than 1, to ensure they remain
unbiased. 
This weighting also increases the noise required for differential privacy,
because the sensitivity of a single observation can now be as high as 
$1/p_{min}$, where $p_{min}$ is the minimum probability of any data
point. 

With these changes, we built a linear model on count-featurized data and
compare it to the (linear) raw-data model used in production. Both models were
trained using VW's online contextual bandit learner; in the production
system, a snapshot of the model is deployed to application servers every five
minutes.

\heading{Applicability.}
Our results suggest that in this application, selectivity is achieved naturally by
retaining only the last day of data in the hot window and without the need for
\sysname's training set minimization.
This is because news is highly non-stationary: new content appears
every hour and breaking news influences people's short-term interests.
As a result, even without \sysname, training models on the last day of raw data
is sufficient, and in fact better than training on more days.
This is in contrast to the MovieLens and Criteo datasets, which
are much more stationary and hence can benefit from Pyramid's training
set reduction.

That said, even in non-stationary settings, \sysname can still enhance data
protection through its privacy-preserving counts.  We compared the estimated
click-through rate (CTR) of the count model (with and without noise) to the raw
model across a seven-day period in April 2016. \F\ref{f:news_progressive_loss} shows the
results relative to the default article ranking by editors. Despite
day-to-day variations, on average count models perform within 7\% and
13.5\% (with noise) of the raw model performance.

\heading{Support for workload evolution.}  We also assessed how 
\sysname would support changes in MSN over time, without 
accessing the raw data store. MSN developers have spent
hundreds (thousands) of human (compute) hours optimizing the production models.
The changes include: tuning hyperparameters
and learning rates, adding L1/L2 regularization, testing different
exploration rates or model deployment intervals, and adding/interacting/removing
features. For example, in some regions regulatory policies prevent certain user
data from being collected, so they are removed and models are retrained.
\sysname supports all of the listed changes (\S\ref{sec:workload-updates})
except adding new features/feature interactions.
\section{Analysis and Limitations}
\label{sec:security-analysis}

We analyze \sysname's security properties in the context of our threat model
(\S\ref{sec:threat-model}), pointing out its limitations.
A \sysname deployment has three components:
(1) A central repository of raw data in cold storage that is infrequently accessed and is assumed to be secure. Protecting this data store is outside of \sysname's scope.
(2) A compute/storage cluster used to train models, store the plaintext hot window, and to store and update count tables.
(3) Numerous model servers storing trained models and cached versions of count tables.

We first examine the effects of compromising the cluster responsible for training models, maintaining the hot window, and storing the count tables.
This will reveal the state of the count tables at time $T_{\textit{attack}}$-$\Delta_{\textit{hot}}$ by subtracting all observations residing in the hot window at $T_\textit{{attack}}$.
Property {\bf P1} in \S\ref{sec:threat-model} captures this exposure.
However, the observations from the range $[T_{\textit{attack}}$-$\Delta_{\textit{retention}}, T_{\textit{attack}}$-$\Delta_\textit{{hot}}]$ are protected through differential privacy (property {\bf P2} in \S\ref{sec:threat-model}).
We expect that the hot window ($\Delta_{\textit{hot}}$) will be small enough that only a small fraction of an organization's data will be exposed.
Observations whose retention period ended before $T_{\textit{attack}}$ will have been erased, and the models will have been retrained to forget this information (property {\bf P3} in \S\ref{sec:threat-model}).

In addition to the hot data, the adversary can siphon observations arriving in the interval $[T_{\textit{attack}}, T_{\textit{attack}}^{\textit{end}}]$.
Hence, the amount of data exposed depends on the time to discover and respond to an attack.
The sliding nature of \sysname's hot window gives the organization an advantage when investigating breaches.
If an organization knows $T_{\textit{attack}}$ and $T_{\textit{attack}}^{\textit{stop}}$, it will be able to determine exactly which observations were exposed to the attacker and take the appropriate steps.
Knowing these times is only required for post-attack auditing, not for protection of past data during the attack.

Under our current threat model, \sysname does not protect data from multiple intrusions happening during the same time window.
If an attacker accesses \sysname's internal count tables, that attack is eradicated, and then gains access again at $T_{\textit{attack2}}$ where $T_{\textit{attack2}}$ follows $T_\textit{{attack}}^{\textit{stop}}$, the attacker will be able to compute the full fidelity count tables for updates that occurred during the time range $[T_{\textit{attack}}^{\textit{stop}},min(T_{\textit{attack2}},T_{\textit{win\_end}})]$ by subtracting the state of the count table at $T_{\textit{attack}}$ from the state of the same count table at $T_{\textit{attack2}}$.
$T_{\textit{win\_end}}$ is the time when \sysname finishes populating the count table it was populating at $T_{\textit{attack}}^{\textit{stop}}$.
One approach to mitigate this attack is to require that \sysname recomputes
count tables after $T_{\textit{attack}}^{\textit{stop}}$,
including reinitializing them with new draws from the Laplacian distribution.
This will require an increased privacy budget but will still provide a privacy guarantee.

\S\ref{sec:evaluation} demonstrates the need to cache count tables on the application model servers. 
Attackers that compromise an application server will gain access to the existing cached count table, trained models, and a stream of plaintext prediction requests (unlabeled observations).
With access only to the application server the adversary will be able to calculate the difference between the existing count table and new count tables as they are replicated.
The adversary will learn little because the difference between the cached count table and the newly replicated count table will be differentially private.

A key limitation of our system stems from our design choice to expose data for a period of time, while it is hot.
Data is exposed through the hot data store, trained models, external predictions, and other states that may persist after the data is phased out into the differentially private count tables.
There are three implications of this design choice.
First, an adversary may monitor these states {\em before} actually mounting the full-system break-in that \sysname is designed to protect against (so before $T_{\textit{start}}$).
\S\ref{sec:threat-model} explicitly leaves this attack out of scope.
Second, exposing the hot data in raw form to programmers and applications may produce data residues that persist after the data is phased out, potentially revealing past information when an attacker breaks in at $T_{\textit{start}}$.
For example, a programmer may create a local copy of the hot window at time $\textit{T}$ for experimentation purposes.
While we cannot ensure that state created out-of-band is securely managed, the \sysname design strives to eliminate any residues for state that \sysname manages.
This is why we enforce model retraining whenever the hot window is rolled over.
And this is why we clarify in \S\ref{sec:differential-privacy} that the count and weight selection mechanisms should incorporate differential privacy.
Third, while the exposed hot data may be small (e.g., 1\% of all the data), it may still reveal sufficient sensitive information to satisfy the attacker's goal.
Despite these caveats, we believe that our design decision to expose a little hot data affords important practical benefits that would be difficult to achieve with a fully protected design.
For example, unlike fully differentially private designs~\cite{mcsherry2009differentially}, our scheme allows training of {\em unchanged} ML algorithms with limited impact on their accuracy. 
Unlike encrypted databases~\cite{popa2011cryptdb,tu2013processing}, our scheme
provides performance and scalability close to---or even better than---running
on the raw, fully exposed data.

\section{Related Work}
\label{sec:relwork}

\headingg{Closest works.}
Closest to our work are the building blocks we leverage for \sysname's
selective data protection architecture:
count featurization and differential privacy.
{\em Count featurization} has been developed and adopted to improve
performance and scalability of certain learning systems.
We are the first to retrofit it to improve data protection,
defining the protection guarantees that can be achieved and
implementing them without sacrificing accuracy.

To implement these guarantees, we leverage {\em differential privacy
theory}~\cite{dwork2006differential}.
The typical threat model for differentially private systems~\cite{Mcsherry:pinq,
roy2010airavat, mcsherry2009differentially} is different from ours: they protect
user privacy in the results of a publicly released computation, whereas
\sysname aims to protect data inside the system, by minimizing access to
historical data so its accesses can be controlled and monitored more
tightly.
For example, differential privacy frameworks (e.g., PINQ~\cite{Mcsherry:pinq}
and Airavat~\cite{roy2010airavat}, adding privacy to LINQ and MapReduce
respectively) ensure that the result of a query will be differentially private.
However, these systems require full and permanent access to the data.
The same holds for privacy-preserving recommender systems~\cite{mcsherry2009differentially}.
Pan-privacy~\cite{dwork2010differential, Chan:2011:PCR:2043621.2043626,
mir2011pan} is a variant of differential privacy that holds even
when an adversary can observe the system's internal state, a threat model close
to ours.

\sysname is the first to combine count featurization
with differential privacy for protection.\footnote{Azure applies tiny levels
of Laplacian noise to count featurization to avoid overfitting, but such low
levels neither provide protection nor raise the challenges we encountered.}
This raises significant challenges at scale, including
rampant noise with large numbers of count tables and damaging interference of
differential privacy noise with count-min sketches.
To address these challenges, our design includes two techniques: noise weighting
and private count-median sketches.
Prior art, such as iReduct~\cite{xiao2011ireduct} or GUPT~\cite{mohan2012gupt}, included a noise weighting scheme to allocate less of the privacy budget to queries with larger results.
To our knowledge, we are the first to point out the limitations of CMS
integration with differential privacy and propose private count-median sketches
as a solution.

\heading{Alternative protection approaches.}
Many alternative protection models exist.
First, many companies enforce a {\em data retention period}.
However, because of the data's perceived benefit, most companies
configure long periods.
Google maintains data for 9-18 months~\cite{andersonGoogleRetention2010}.
\sysname limits the data's exposure for as long as the company decides to
retain it.
Second, some companies {\em anonymize data}: Google
erases the last byte of IP addresses in search logs after 6
months~\cite{googlePublishPolicyEuropoeanCommision2008}.
Anonymization provides very weak protection~\cite{Narayanan:2008:RDL:1397759.1398064}.
\sysname leverages differential privacy to provide rigorous protection
guarantees.
Third, some companies enforce {\em access controls} on the data.
Google's Sawmill strips out sensitive data before returning results to
processes lacking certain permissions~\cite{beckerReplacingSawzall2015}.
Given the push toward increased developer access to data~\cite{google-data-policy, hearst-data}, \sysname provides additional
benefit by protecting data on a needs basis.

\heading{Data minimization.} Compact data representation is an important
topic in big data systems, and many techniques exist for different scenarios.
{\em Sketching techniques} compute compact representations of the data that
support queries of summary statistics~\cite{cormode2005cms}, large-scale
regression analysis~\cite{mahoney2011randomized},
privacy preserving aggregation~\cite{melis2015efficient};
{\em streaming/online algorithms}~\cite{muthukrishnan2005data,shalev2011online}
process the data using bounded memory, retaining only the information relevant
for the problem at hand;
{\em dimensionality reduction techniques}~\cite{burges2010dimension} find a
low-dimensional, faithful representation of the raw data, according to
different measures of faithfulness;
{\em hash featurization}~\cite{shi2009hash} compacts high-cardinality
categorical variables;
{\em coresets}~\cite{feldman2009private,agarwal2005geometric} are data subsets giving a good approximation for a given computation;
{\em autoencoders} attempt to learn a compressed identity function~\cite{GoodfellowDeepLearningBook}.

We believe that this rich literature should be inspected for candidates for
selective data protection.  Not all mechanisms will be suitable.
For example, according to our evaluation (\F\ref{f:counts_vs_raw}), hash
featurization~\cite{shi2009hash} does not yield sufficient training set
reduction.
And none of the mechanisms listed above appear to support workload evolution.
The next section presents a few promising techniques we have identified.
\section{Closing: A Vision for Selectivity}
\label{sec:selectivity}

We close with our vision for selectivity in big data systems.
Today's indiscriminate data collection, long-term archival,
and wide-access practices are risky and unsustainable.
It is time for a more rigorous and selective approach to big data collection,
access, and protection so that its benefits can be reaped without undue risks.

\begin{wrapfigure}{R}{0.22\textwidth}
  \centering
  \footnotesize
  \vspace{-0.5cm}
  \includegraphics[width=\linewidth]{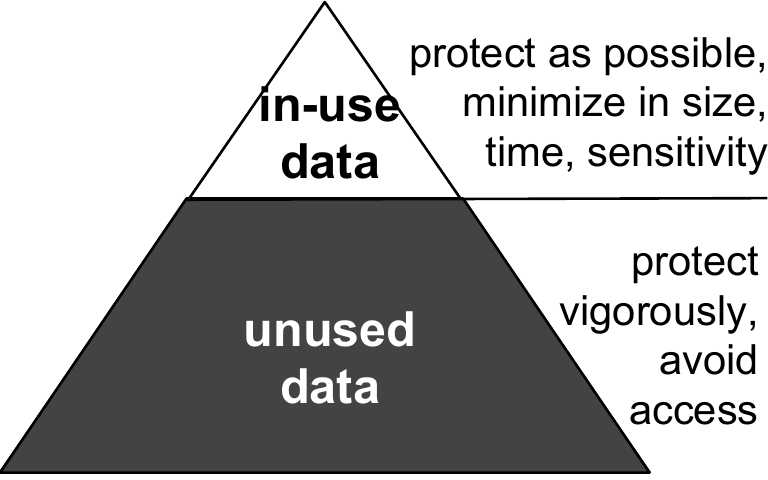}
  \vspace{-0.7cm}
\end{wrapfigure}
Our vision (illustrated on the right) involves architecting data-driven
systems to permit clean separation of data needed by current and
evolving workloads, from data collected and archived for possible future needs.
The former should be minimized in size and time span (hence the pyramid
shape).
The latter should be protected vigorously and only tapped under exceptional
circumstances.
These requirements should be met without disrupting functional properties of
the workloads.

The notion of selectivity applies to many big data workloads, including
ML and non-ML, and there are perhaps multiple ways to conceptualize the
data selectivity problem.
For ML workloads, we find that a productive way of identifying potential mechanisms
is to model the problem as a {\em training set minimization problem}.
This reveals a rich set of mechanisms that might be leveraged to achieve data selectivity.
We have identified several promising mechanisms, which we hope to
incorporate into \sysname for wider workload coverage:

\noindent
$\bullet$
{\em Vector quantization (VQ).}
VQ~\cite{gersho2012vector} is a family of techniques used
to compactly represent high dimensional, real-valued feature vectors.
At a high level, VQ computes a small subset of vectors, known as the
codebook or the centroids, that are representative of the entire set of input
vectors (e.g., historical data).

\noindent
$\bullet$
{\em Sampling.} Uniform random sampling and more advanced techniques like
herding~\cite{chen2010super} can be used to maintain a representative sample of
the historical data. This sample can be combined with in-use data to form a
training set. Compared to VQ, which often makes certain assumptions about
the underlying data (\eg that it forms clusters), sampling techniques are more
general.

\noindent
$\bullet$ {\em Active learning.}
Active learning algorithms~\cite{settles2012active} tell users what specific
data points they need for improved accuracy.
Originally built to decrease manual labeling, they may be valuable to
selective data collection.

We leave investigation of such mechanisms for future work.
The key challenge will be to identify the kinds of protection and privacy guarantees
achievable with these mechanisms, and how to effectively implement them.
This paper provides a first blueprint for this process.

\section{Acknowledgements}

We thank our shepherd, Ilya Mironov, and the anonymous reviewers for
their valuable feedback.
We thank Alekh Agarwal, Markus Cozowicz, Daniel Hsu, Angelos Keromytis, Yoshi
Kohno, John Langford, and Eugene Wu for their feedback and advice.
This work was supported in part by NSF grants \#CNS-1351089 and \#CNS-1514437, a Sloan Faculty Fellowship, a Microsoft Faculty Fellowship, and a Google Ph.D. Fellowship.
 \appendix
\begin{figure*}[ht!]
\subfigure[{\bf MovieLens private noise weighting}]{
  \includegraphics[width=0.31\linewidth]{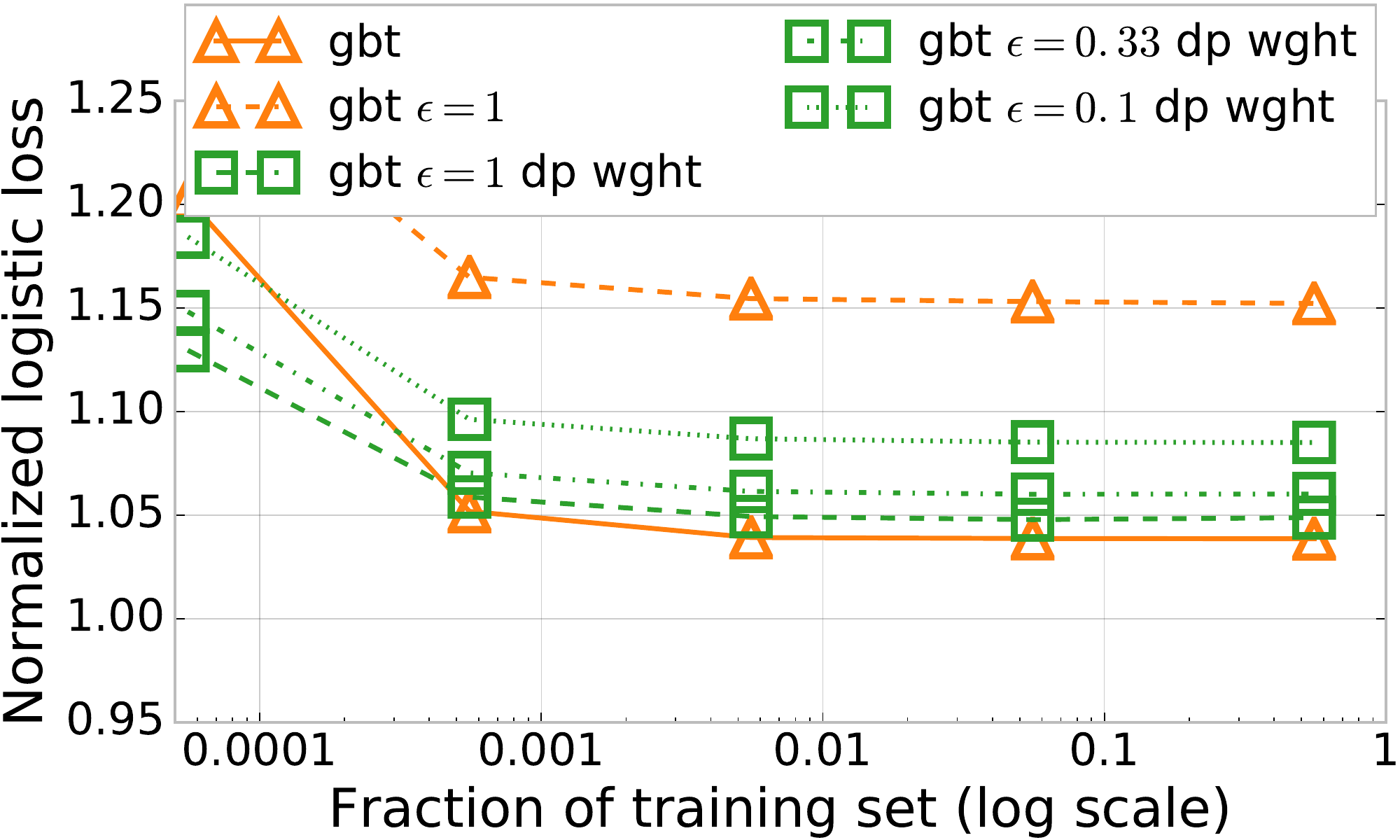}
  \label{f:movielens_private_weighting}
}
\subfigure[{\bf Criteo-Kaggle nn private noise weighting}]{
  \includegraphics[width=0.31\linewidth]{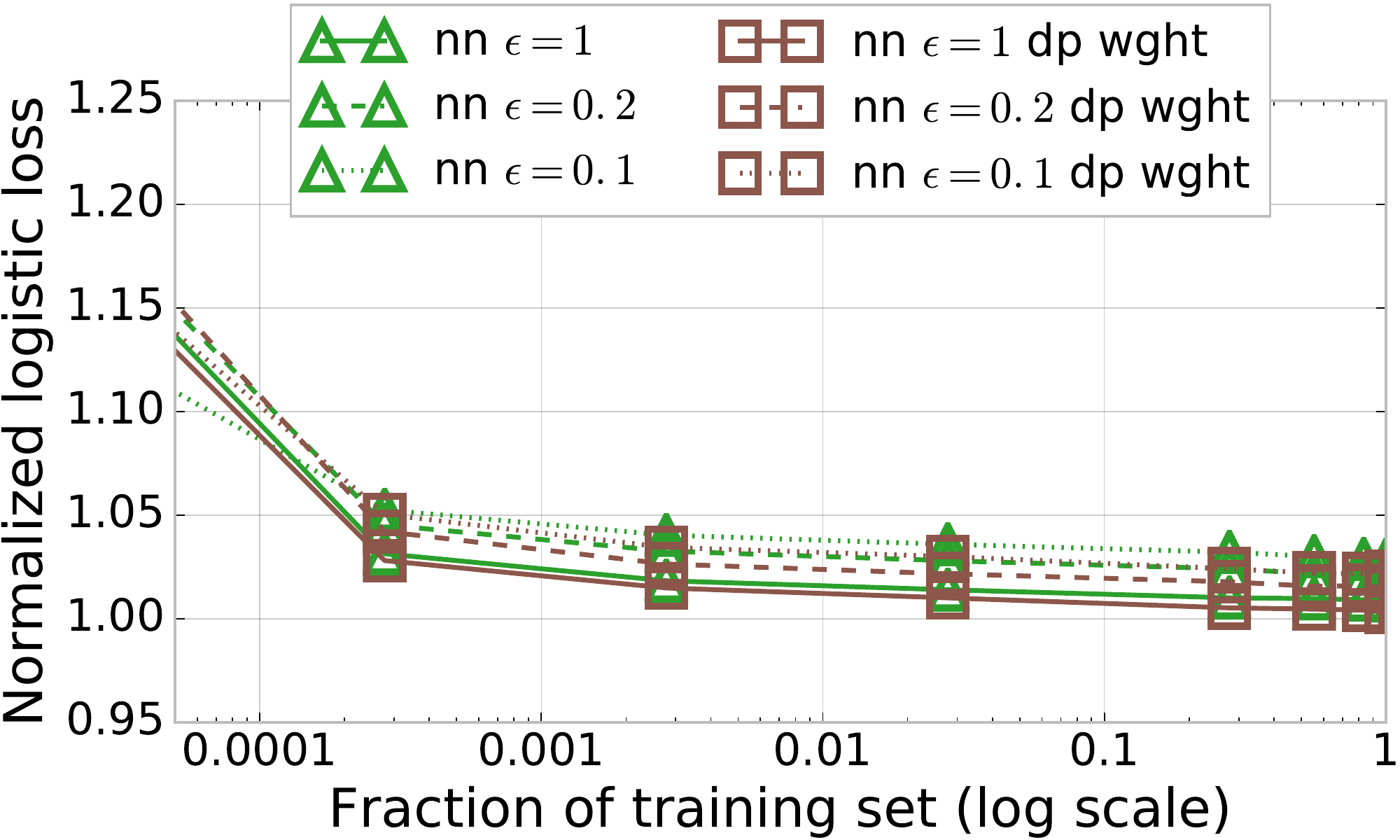}
  \label{f:criteo_nn_private_weighting}
}
\subfigure[{\bf Criteo-Kaggle logistic regression private noise weighting}]{
  \includegraphics[width=0.31\linewidth]{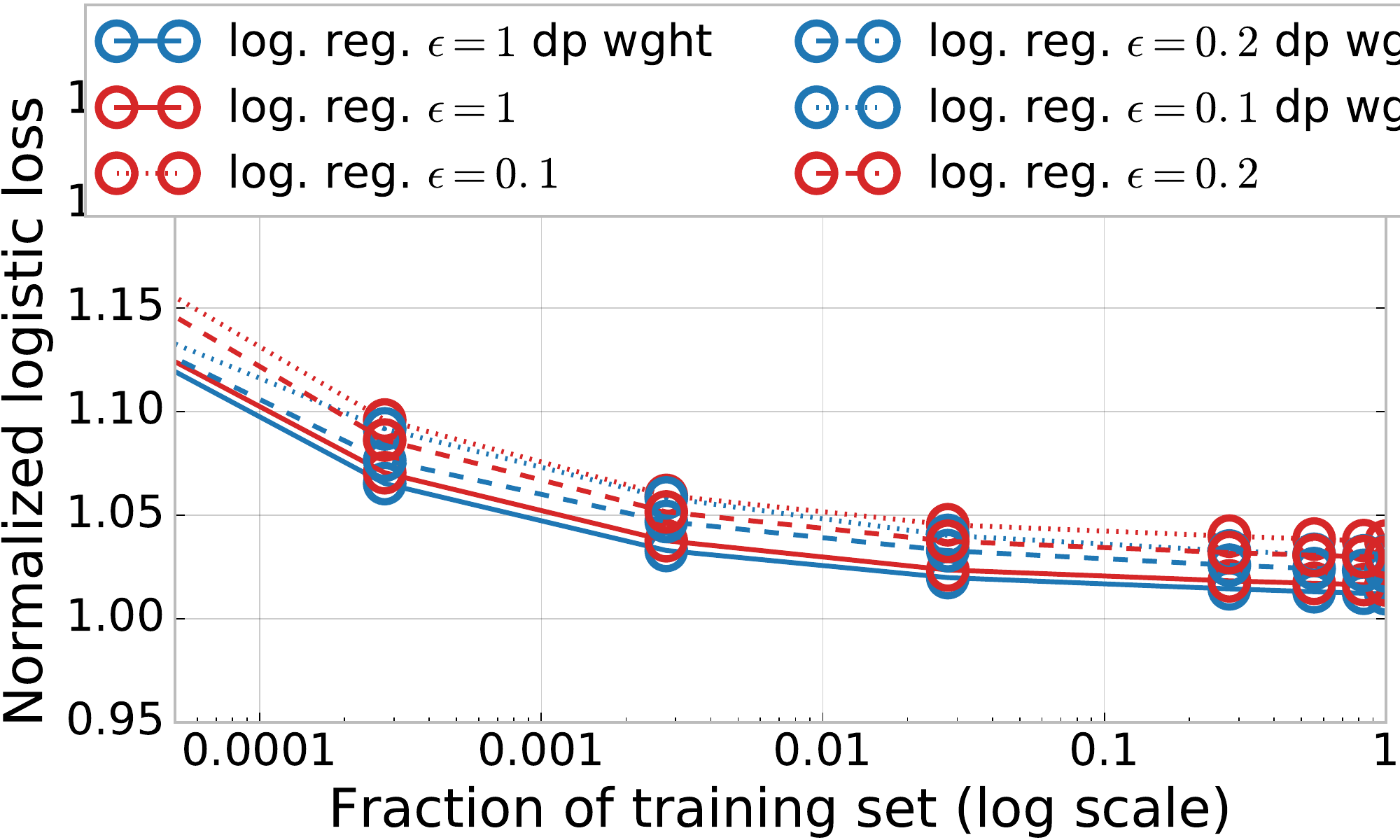}
  \label{f:criteo_linear_private_weighting}
}
\vspace{-0.35cm}
  \caption{\footnotesize {\bf Private Weighted Noise Infusion}
      Results are normalized to the respective baselines.
      Weights were calculated on a 200K observation window for MovieLens and a 1M observation MovieLens for Criteo-Kaggle.
      \sysname can provide privacy to the observations used to calculate the hot window while still effectively distributing the privacy budget across features.
  }
\label{f:dp_weighting}
\vspace{-0.3cm}
\end{figure*}

\subsection{Differentially Private Weight and Count Selection}
\label{s:dp-processes}

As noted in \S\ref{sec:differential-privacy}, the weighted noise infusion and count selection processes must be made differentially private.
While our IEEE Security \& Privacy paper~\cite{pyramid-sp17} did not address this problem, we have since modified \sysname to compute feature weights in a differentially private way.  We also have a design for private count selection.
Our method is based on several known techniques from the differentially privacy literature (overviewed in \S\ref{s:smooth-sensitivity}), which we adapt to our specific problem.
This section describes and evaluates our mechanism for private weight computation
(\S\ref{s:dp-weights}) and describes how one might apply the same mechanism to make count selection private
(\S\ref{s:dp-groups}).

\subsubsection{Background: Smooth Sensitivity}
\label{s:smooth-sensitivity}

Smooth sensitivity~\cite{Nissim:2007:SSS:1250790.1250803} is a technique used to fine-tune the amount of differential privacy noise to the sensitivity of a computation on a specific dataset, instead of the worst case sensitivity, which in many cases can be disastrous.
Smooth sensitivity is based on the insight that for some statistics the worst case sensitivity is very large (e.g. the whole range of the data for the median), but on most datasets changing a single data point barely changes the result, resulting in a small local sensitivity.
One can add noise based on the smooth sensitivity, an upper bound of the local sensitivity that prevents the local sensitivity to leak any information on the dataset.

For some functions $f$, computing the smooth sensitivity with a closed formula is not practical, or even not possible.
In such cases, and assuming $f$ can be approximated well on subsets of the data, it is possible to leverage the sample-and-aggregate framework~\cite{Nissim:2007:SSS:1250790.1250803}.
One splits the full database into $n$ groups (as in \cite{Smith:2011:PSE:1993636.1993743}) and applies function $f$ to each of the groups.
The results are then aggregated using a function with a known smooth sensitivity, such as the median or the center-of-attention, and adds noise to the output of the aggregation function.
Since each data point can change at most one of the groups, the final result is differentially private.

\subsubsection{Private Weighted Noise Infusion}
\label{s:dp-weights}

We refine the weighted noise infusion process described \S\ref{sec:differential-privacy}) to make it differentially private using smooth sensitivity.
The quantile function indeed has a poor global sensitivity but on most datasets, including those we tried, has a very small local sensitivity.
We adapt the J-List algorithm for median smooth sensitivity~\cite{Nissim:2007:SSS:1250790.1250803}, modified in a straight forward way for arbitrary quantiles, to compute the smooth sensitivity of the quantile function for each feature over one observation window.
The maximum value of each count is the size of the window, and we use a Cauchy distribution to preserve $(\epsilon)$-differential privacy.

We evaluate our private weighted noise infusion process with the same datasets and a similar experimental setup as in our evaluation (\S\ref{s:methodology}).
For simplicity, for each dataset, we set aside a window to be only for weight calculation and are not reused for training later.
We choose a window size of 1M points, small compared to the size of the datasets (2.5\% of the dataset for Criteo, and 4.5\% for Movielens -- although results a identical with 200k windows, less than 1\% of the data, on the Movielens dataset) but large enough to get reliable differentially private estimates.
We compute the $10^{th}$ percentile of the counts as it is less sensitive than the $1^{st}$ one, and then rescale the weights to be between $1$ and $10k$, the maximum number used in the non private weights computation (\S\ref{sec:evaluation}).
We compute the noise weights on the first window using the entire privacy budget, and then use the results to initialize the count tables for training on the rest of the training set.
We use the same $\epsilon$ value for both weight calculation and training.

\F\ref{f:dp_weighting} shows our results.
For both datasets, the private weighted noise infusion preserves the performance gains we observed for the non-private weighted noise infusion.
On the Criteo dataset, we observe the same improvement as with non private noise weighting, about 0.5 percentage points. On the Movielens data, the results are even better, which we assume is due to the larger window used to compute the weights. We see that with $\epsilon=1$ the weighting scheme allows the boosted trees to get within 5\% of the baseline. The improvements for $\epsilon=0.33$ and $\epsilon=0.1$ are even larger, with $\epsilon=0.33$ close to the 5\% bar.

\subsubsection{Private Count Selection}
\label{s:dp-groups}

Like the initial design for weighted noise infusion, our current group selection mechanism is not differentially private and will leak information about the data used in performing count selection.
While we have not done so yet, we think that a good approach to make differentially private count selection is to leverage the sample-aggregate framework from~\cite{Nissim:2007:SSS:1250790.1250803}, with the center of attention aggregation.
The center of attention can aggregate functions with multi-dimensional outputs. This means we can compute our conditional mutual information metrics for each group we are interested in, and add noise proportional to the smooth sensitivity of an aggregation on the entire vector.
We can then chose the groups with the higher results as before, thus preserving privacy.

\subsubsection{Extension}

A promising extension for our private weight infusion mechanism would be to refine the count estimates over multiple hot windows, yielding a double improvement: (1) counts are more accurate, as they benefit from the previous weighting, and (2) previous weights can also be used to compute the new ones more accurately.
When weight estimates are computed on multiple windows, all the information can still be used to get more precise estimates.
As explained in \cite{Xiao:2011:IDP:1989323.1989348}, the lowest-variance, unbiased estimate of the weights from multiple computations is $w = \sum_{i=1}^{n}\frac{w_i}{\lambda_i^2} / \sum_{i=1}^{n}\frac{1}{\lambda_i^2}$, with $w_i$ the weight computed on window $i$ with noise scale $\lambda_i$.
For instance, one could choose to assign half the privacy budget of $n$ consecutive windows to computing the weights.
Counts are computed without weighting on the first window. For each of the following $n-1$ windows, weights are computed using the previous' window counts, and half the current privacy budget. These new values are merged with any previous estimate as we just described.
The counts for this window are then computed with the other half of the privacy budget and the new weights.
This process can be repeated regularly, every month for example, to update the weights to changes in the distribution.


\subsection{Avenue for Deployment: \sysname As-A-Featurization-Service}
\label{s:featurization_service}

We discuss a possible approach to deployment in a production setting.
As described in this paper, \sysname requires substantial changes to a company's data pipeline and proposes replacing the past with summary statistics.  The accuracy loss, while reasonable, may be unacceptable to some.
Still, we believe that there are avenues for \sysname's immediate adoption in production.
One of those, which we are presently investigating, is to provide \sysname's differentially private count featurization as a service.

Count features or marginals are commonly used in addition to raw features to improve the performance of machine learning models that are trained on full datasets.
As demonstrated in \S\ref{sec:performance_eval}, it is critical for such applications to collocate the count tables with the predictive models.
However, widely distributing the count tables to model servers increases their exposure.
We argue that using \sysname's approach in such applications would already produce a large benefit in reduced data exposure (for instance in the case of a compromised model server), and we believe the accuracy cost would be small enough to be bearable.
This way, an organization does not need to replace the entire data management system to benefit from some techniques leveraged by \sysname when deploying count featurization.

To retrofit \sysname to this use case, we advocate building a count featurization service. This service would plug in the data injection pipeline, and process all incoming data to build the differentially private count tables. The count tables would use the count-median sketch for high cardinality features, and could support $(\epsilon,\delta)$-differential privacy and advanced composition to support more features without too much impact.
It would divide the data stream in windows of a parametrized size or time, and support retention policies.
Noise weights could automatically be computed and refined over time, as described in \S\ref{s:dp-weights}. Automatic feature groups could be supported if made differentially private.
The featurization service would provide a client library that can be called to add the count features to datapoints before they are used for training or prediction. This library would be responsible for caching the count tables, allow low latency featurization necessary in many settings.

\bibliographystyle{IEEEtran}
\bibliography{abbrev,conferences,refs,my_papers}

\end{document}